\documentclass[superscriptaddress,pra,aps,twocolumn,10pt,longbibliography]{revtex4-1}

\usepackage{amsfonts}
\usepackage{amssymb}
\usepackage{amsmath}
\usepackage{graphicx}
\usepackage{verbatim}
\usepackage{hyperref}
\usepackage[english]{babel}
\usepackage{natbib}
\usepackage[pdftex]{color}
\usepackage[dvipsnames]{xcolor}

\hypersetup{
    colorlinks=true,       
    linkcolor=cyan,          
    citecolor=magenta,        
    filecolor=magenta,      
    urlcolor=cyan,           
    runcolor=cyan
}

\graphicspath{{figs/}}

\def \l {\left}
\def \r {\right}

\newcommand{\bra}[1]{\langle#1|}
\newcommand{\ket}[1]{|#1\rangle}
\newcommand{\braket}[2]{\langle#1|#2\rangle}
\newcommand{\ketbra}[2]{|#1\rangle \langle #2 |}
\newcommand{\ketbrad}[1]{|#1\rangle\!\langle #1|}
\newcommand{\ketbradt}[2]{|#1\rangle\!\langle #2|}
\newcommand{\mean}[1]{\langle#1\rangle}

\newcommand{\citedfn}[1]{\setcitestyle{open={[[},close={]]}}\cite{#1}\setcitestyle{square}\hspace{-0.125cm}}

\begin{document}
\title{Incoherent Approximation of Leakage in Quantum Error Correction}

\author{Jeffrey Marshall}
\affiliation{QuAIL, NASA Ames Research Center, Moffett Field, California 94035, USA}
\affiliation{USRA Research Institute for Advanced Computer Science, Mountain View, California 94043, USA}

\author{Dvir Kafri}
\affiliation{Google Quantum AI, Santa Barbara, CA 93117, USA}

\begin{abstract}
 Quantum error correcting codes typically do not account for quantum state transitions - leakage - out of the computational subspace. Since these errors can last for multiple detection rounds they can significantly contribute to logical errors. It is therefore important to understand how to numerically model them efficiently. Fully quantum simulations of leakage require more levels per leaked qubit, which substantially limits the system sizes that may be simulated. To address this, we introduce a Subspace Twirling Approximation (STA) on quantum channels that preserves the incoherence between the computational and leakage subspaces. The assumption of incoherence enables the quantum simulation of leakage at little computational overhead. We motivate the approximation's validity by showing that incoherence is achieved naturally during repeated stabilizer measurements. Additionally, we provide various simulation results which show that the STA yields accurate error correction statistics in the repetition and surface codes with physical error parameters.

\end{abstract}

\maketitle

\section{Introduction}

Quantum error correction provides a means for implementing quantum computation in a fault-tolerant manner, despite the presence of unavoidable physical noise \cite{knill-laflamme}. This is demonstrated by threshold theorems, which show that as long as the physical error rate is below some threshold value, logical errors can be exponentially suppressed in the size (distance)  of a code \cite{threshold_1, threshold-non-markov, threshold-long-range}. This then guarantees that, with high probability, logical information can be stored and manipulated for times sufficient for useful quantum computation. 
The threshold theorems all make certain assumptions about the noise, and when these are violated, often a sharp decrease in code performance results. 
This is particularly true for non-computational errors, where degrees of freedom (physical qubits) transition to physical states outside of the (typically two-level) computational subspace \cite{fowler-leakage-topological, google-leakage-2022, google-removing-leakage-2021, leakage-mitigation}. 

Non-computational errors occur in most proposed quantum computing architectures in some fashion. This can range from loss in photonic \cite{LOQC-overview, Terhal_2020} and trapped ion settings \cite{trapped-ion-leakage}, to excitations out of the qubit subspace in superconducting qubit architectures \cite{google-leakage-2016, Gambetta-leakage, google-leakage-2022}.
Superconducting qubits have been a promising candidate for constructing large scale programmable quantum chips, with the earliest demonstrations of a quantum over classical advantage \cite{google-q-suprem}, as well as encouraging results for scaling of a surface code \cite{scaling_SC_Google_2022}. 
A superconducting qubit is typically modeled as a quantum anharmonic oscillator, where the lowest two (qubit) levels span the computational subspace and excited (``leaked'') states are non-computational. The process of exciting states outside of the computational subspace is known as ``leakage'' \cite{wood2018quantification}. 
Though great care is taken to avoid such transitions, they are inevitable with current technology, and can arise from simple heating mechanisms or more complex coherent transitions \cite{simple-pulses, reduced-phase-error, understanding-leakage, chen_measuring_2016, lambda-leakage}. Notably,
leakage errors can greatly reduce code performance of most error correcting codes \cite{fowler-leakage-topological, handling-leakage-subsys-code, leakage-detection}. As such, many proposals exist to try to handle or remove leakage as effectively as possible in experimental realizations of error correction \cite{google-removing-leakage-2021, leakage-mitigation, handling-leakage-subsys-code, critical-faults-leakage, Terhal-2021-LRU, understanding-leakage, google-leakage-2022}.

In addition to posing a challenge to error correction, leakage also dramatically increases the cost of simulating such experiments. 
While there do exist efficient classical simulators of many error correcting codes relying on the stabilizer representation \cite{gidney2021stim, aaronson-gottesman-stabilizer} -- which can accommodate thousands of qubits -- certain noise models (such as leakage) do not fit naturally into this framework. This motivates the need for fully quantum simulations for use in verification and validation.
However, a fully quantum simulation including leakage becomes severely limited in the system sizes that can be addressed at reasonable computational cost.
For example, the memory requirements for such a simulation of $n$ three-level systems scales as $O(3^n)$ compared to $O(2^n)$ in the qubit case. 
Given a fixed budget, this effectively reduces the achievable system size $n$. As an example, the distance 5 rotated surface code \cite{bombin-rotated-SC, Horsman_2012} involves 25 data qubits, which for qutrits would be equivalent to a 40 qubit simulation.
In this paper we discuss a general method for approximating leakage as an incoherent process. The incoherence of transitions between the computational and leaked subspaces means that a given qubit will never be in superposition of more than two basis states.
This allows for quantum simulations with leakage (3 or 4 levels per qudit), but only using the resources of a qubit (2 level) system. The approximation can be derived as a Subspace Twirling Approximation (STA) at the level of noisy quantum channels [Sect.~\ref{sec:STA}].

While the STA, in principle, can be applied to any quantum circuit, it is particularly amenable to error correction circuits with repetitive stabilizer measurements. Such measurements cause the coherence between the qubit and leakage subspace to decay exponentially in the number of error correction rounds [Sec.~\ref{sec:coherence-decay}].
For realistic models and parameters based on recent experiments \cite{scaling_SC_Google_2022}, the STA faithfully reproduces the full qutrit simulation statistics in repetition (up to distance 9) and surface codes (distance 3) [App.~\ref{sec:STA-physical}].
In the case where the leakage mechanism is purely coherent, using the STA results in a discrepancy with exact simulation. Even so, we show that one can derive an effective thermal leakage model which is both consistent with coherent leakage and for which the STA is an accurate approximation. The effective model accurately predicts the ideal simulation statistics, such as leakage populations and logical error rates [Sect.~\ref{sec:STA-justification}]. This is particularly useful, as we require leakage to be incoherent to efficiently simulate large scale error correction experiments in the stabilizer formalism \cite{scaling_SC_Google_2022}.

\subsection{Basic discussion of leakage origins}

In order to motivate the study of leakage we focus our attention on weakly nonlinear superconducting qubits (transmons) \cite{koch_charge-insensitive_2007}. The bulk of our analysis will be agnostic to the precise system and, in particular, it also readily applies to flux qubits \cite{c-shunt-flux-qubit}. We do not consider cases when a qubit is lost from the system entirely (such as can occur in photonic or trapped ions), though we note that including classical degrees of freedom (discussed in the next section) should enable proper modeling of these effects as well. 
The low-energy sector of a single qubit's Hamiltonian is described by
\begin{equation}
    H / h = f_{10} n - \frac{\eta}{2}n(n-1),
    \label{eq:transmon}
\end{equation}
where $n = a^\dagger a$ represents a number operator counting excitations in the circuit. The frequency $f_{ij}$ sets the energy splitting between the standard basis states $\ket{i}$ and $\ket{j}$, while the nonlinearity parameter is set by $\eta=f_{10}-f_{21}$. Typical values for transmon qubits are  $f_{10}\sim$4-6 GHz, $\eta \sim$200-300 MHz, while for C-shunt flux qubits $f_{10}$ is in the same range, with $-\eta \sim$500-900MHz having opposite sign \cite{c-shunt-flux-qubit}.
Single qubit gates are usually implemented using microwave drives, which typically appear as time dependent offset terms proportional to the charge operator $i(a^\dag-a)/\sqrt{2}$. Leakage in this context refers to any excitation of a qubit beyond the first energy state $\ket{1}$. 

There are several documented physical mechanisms causing leakage in transmons. Single qubit microwave gates can cause coherent transitions between $\ket{1}$ and $\ket{2}$, due to finite spectral components in the drive near frequency $f_{21}=f_{10}-\eta$. Usually these gates are calibrated to remove these side-bands \cite{motzoi_simple_2009,gambetta_analytic_2011}, but control crosstalk between drive lines targeting different qubits can enhance these effects. Similarly, control error may also cause leakage during two-qubit gates. The details of this leakage depend on the gate implementation \cite{tunable-coupler-Yan,cr-gate-Paraoanu,cr-gate-rigetti,parametric-gate-Chow}, but typically this is observed as a coherent transition between two-qubit states (e.g., $\ket{11}\leftrightarrow\ket{02}$). Leakage is also caused by incoherent processes such as interaction with a passive environment. This is particularly relevant to ``diabatic" CZ gates \cite{foxen_demonstrating_2020,sung_realization_2021,werninghaus_leakage_2021}, where  state $\ket{2}$ has high occupation during the gate. In practice, there are several other sources of leakage in superconducting qubits, ranging from passive ``heating'' \cite{chen_measuring_2016}, transport of excitations mediated by undesired resonances during the gates \cite{google-leakage-2022}, or measurement-induced state transitions \cite{sank2016measurement, khezri2022measurementinduced}.

\section{Open system simulations of error correction codes with leakage}

\subsection{Trajectories simulations of quantum channels}
\label{sec:traj-sims}

Here we discuss the simulation of noisy quantum circuits using quantum trajectories.
In the context of the present work, an idealized quantum circuit simply consists of local unitary operations (gates), measurements, and reset operations (relevant to error correction, where reset effectively cools a qubit to the $\ket{0}$ state).
Noise can be introduced in such a simulation at various levels, e.g. as modifications to the unitary gates, imperfect measurements/reset, or through dedicated noise operations (described below), such as for decoherence. Though the presentation below will be general, in our simulations we restrict ourselves to Markovian models of noise.

A quantum operation can be described mathematically as a completely-positive and trace preserving (CPTP) map acting on the system's density matrix \cite{nielsen_chuang_2010}. Such maps are often known simply as `quantum channels'.
A quantum channel can be represented through its action on a density matrix $\rho$,
\[
\mathcal{\mathcal{E}}(\rho)=\sum_{j\geq1}K_{j}\rho K_{j}^{\dagger}\,.
\]
The set of operators $K_{j}$ are called the ``Kraus operators'', which to preserve the trace of $\rho$  must satisfy the identity $\sum_{j\geq1}K_{j}^{\dagger}K_{j}=I$ \cite{nielsen_chuang_2010}. A noisy quantum circuit can then be realized as a composition of such channels, {$\mathcal{E} = \mathcal{E}_n \circ \dots \circ \mathcal{E}_1$}.
In our simulations, we typically represent a single noisy operation (such as evolution during a gate) $\mathcal{E}_G$, as a composition $\mathcal{E}_G=\mathcal{E}'\circ \mathcal{E}_U$ of $\mathcal{E}_U$. Here $\mathcal E_U$ denotes the ideal target unitary (single Kraus operator $K_1=U$) and $\mathcal{E}'$ takes into account the relevant error mechanism (such as dissipation or leakage). In the limit where multiple weak error mechanisms are present, we can compose multiple channels $\mathcal E'$ after $\mathcal E_U$. For a system with exponentially large dimension $N$ (e.g. $N=2^n$ for $n$ qubits), a naive simulation
of the channel on an arbitrary initial density matrix $\rho$ would require
multiple matrix-matrix multiplications. Since the matrices above have
dimension $N\times N$, the space required will scale at least as
$O(N^{2})$ and the total number of operations will scale as $O(N^{3})$ (the exponent can be reduced slightly using certain optimized matrix algorithms \cite{duan2023faster}).
Fortunately, it is possible to improve on these naive estimates, in both the memory and time overhead, at the expense of error due to sampling.

The quantum trajectories method is a technique for sampling the output
of a sequence of quantum channels, which, without further assumptions, scales as $O(N^{2})$ (instead
of $O(N^{3})$) \cite{trajectories_barenco_1997}. While originally used to describe continuous-time
open system dynamics  (at the level of master equations and Lindblad
operators \cite{carmichael-open-QO}), it can be adapted to discrete time evolution (at the
level of quantum channels and Kraus operators). 
In the discrete case, a single trajectory corresponds to probabilistically applying a sequence of Kraus operators to the initial state. Assuming the initial state is pure, and that all Kraus operators have bounded locality (i.e., act only on $O(1)$ qubits), the memory and time requirement to sample a single trajectory can be reduced to $O(N)$.
This scaling is essential to allow for simulations of a distance 5 rotated surface code \cite{bombin-rotated-SC, Horsman_2012}. Although they provide a quadratic improvement in memory requirements over standard density matrix simulations, trajectories simulations produce more limited information about an experiment, albeit still producing a \emph{faithful sampling} of the measurement statistics.

Each ``trajectory" in the quantum trajectories method represents an independent
experimental simulation of a pure state. To simulate a trajectory, we assume a pure initial quantum
state $\ket{\psi}$. As described above, evolution of a noisy quantum circuit can be described as a sequence
of quantum channels $\mathcal{E}_{1},\mathcal{E}_{2},\,...$, each
composed of Kraus operators $\{K_{j}^{(1)}\}_j,\{K_{j}^{(2)}\}_j, \dots$ Instead of applying a given channel to a density matrix, \emph{one} of its Kraus operators $K_{j}$ is sampled using the Born Rule probability
\[
p(j)=||K_{j}\ket{\psi}||^{2}=\bra{\psi}K_{j}^{\dagger}K_{j}\ket{\psi}.
\]
The result is the normalized state $\frac{1}{\sqrt{p(j)}}K_{j}\ket{\psi}$.
The random sampling and state update is identical to the back-action
of measuring a positive operator-valued measure (POVM) with elements
$M_{j}=K_{j}^{\dagger}K_{j}$. As such, at the end of each trajectory
one is given the record $j_{1},j_{2},...$ of all sampled Kraus operators, including any measurement outcomes. A more detailed account of the quantum trajectories method can be found in Ref.~\cite{google-sims} (e.g. see Alg.~2).

\subsection{Representation of classical and quantum channels}
\label{sec:channel-rep}

One important feature of our trajectories simulation is that the state
vector (and corresponding Hilbert space) can dynamically change size.
Combined with the ``reordering trick'' of Ref.~\cite{obrien-sc-dm-sim}
(see App.~\ref{sec:reordering-trick}), this allows us to simulate a distance 5 rotated surface
code (49 qubits) while representing at most 26 qubits in memory at any time (that is, only one measure qubit is in memory at any instance). A quantum channel that changes the Hilbert space dimension will
have Kraus operators described by non-square matrices. For example,
we can formally describe a quantum operation that creates a new subsystem (degree of freedom). This corresponds to a single Kraus operator $K_0: \mathcal{H}_S \rightarrow \mathcal{H}_S \otimes \mathcal{H}_{S'}$ whose action on a state of system $S$ is 
\[
K_{0}\ket{\psi}_{S}=\ket{\psi}_{S} \otimes \ket{\phi}\,
\]
where $\ket{\phi}$ is the state of the prepared subsystem $S'$. Accepting
a slight abuse of notation, we can express this Kraus operator as
$K_{0}= I_{S}\otimes \ket{\phi}$. Similarly, we can represent a destructive measurement of the same subsystem using Kraus operators
\[
K_{j}= I_{S}\otimes \bra{j} \,.
\]
In both cases, one can immediately verify that the trace-preservation
property $\sum_{j}K_{j}^{\dagger}K_{j}=I$ is satisfied. In practice,
the tensor product structure of each Kraus operator makes it convenient
to represent it numerically as a tensor\citedfn{fn-tensor-indices}.

We can also generalize our quantum channels description to include
classical registers. This generalization (see the Kraus operators in~\eqref{eq:qinstrument} and~\eqref{eq:qinstrument2} below) allows us to model quantum instruments\cite{davies1970operational}. These are useful in contexts where classical information is required to describe the evolution of quantum systems. Examples include quantum communication\cite{chitambar2014everything}, error correction\cite{characterizingquantuminstruments}, and quantum metrology\cite{mclaren2023stochastic, mitra2022compatibility}.

The defining property of a classical register is that, as a local subsystem, it is always \emph{incoherent} with respect to a fixed basis $\{\ket{j}\}$. (For a quantum trajectory, this means that the subsystem always remains in a product state.) While incoherence is
discussed in general detail in Sect.~\ref{sec:simulating_incoherent}, for the case of classical
registers this means that the state vector is always supported by
\emph{exactly one }basis operator. In other words, at any given time
in the trajectories simulation, exactly one observable of the set
$\{\ketbrad{j}\}$ has non-vanishing expectation value. Since
the full system will always be of tensor product form 
$$\ket{\psi}_{S}\otimes\ket{j_{1}}\otimes\ket{j_{2}}\,...$$
the cost of including classical registers is \emph{linear} (instead
of exponential) in the number of classical subsystems. In other words, the observable associated with $\{\ketbrad{j}\}$ is always a ``good quantum number''.

Quantum channels conditioned on classical registers can be expressed in terms of Kraus operators of the form
\begin{equation}
    \label{eq:qinstrument}
    K_{j}^{(r)}\otimes\ketbrad{r},
\end{equation}
where $K_{j}^{(r)}$ acts on the quantum degrees of freedom and $\ketbrad{r}$
on the classical. (For simplicity we let $r$ represent a multi-index of all register values. In practice most channels will only depend on at most one register's value.) One can readily verify that the trace preservation condition is satisfied:
\begin{align*}
 & \sum_{j,r}\l(K_{j}^{(r)}\otimes\ketbrad{r}\r)^{\dagger}K_{j}^{(r)}\otimes\ketbrad{r}\\
= & \sum_{r}\l(\sum_{j}\l(K_{j}^{(r)}\r)^{\dagger}K_{j}^{(r)}\r)\otimes\ketbrad{r}\\
= & \sum_{r}I\otimes\ketbrad{r}=I.
\end{align*}
These operations act by applying the quantum
channel $\{K_j^{(r)}\}_j$ to the quantum degrees of freedom, conditioned on the current unique value of the classical registers in state $r$. Importantly, in order to preserve incoherence between classical registers, Kraus operators above must always be diagonal in the fixed basis of each register. 

Like quantum degrees of freedom, we allow classical registers to also be created or destroyed. The Kraus operators representing these processes take the same form as in the quantum case described above. One important generalization is Kraus operators of the form
\begin{equation}
    \label{eq:qinstrument2}
    K_j \otimes \ket{j} 
\end{equation}
where $K_j$ acts on the quantum degrees of freedom and $\ket{j}$ denotes the value of a new register. These Kraus operators can be used to model the \emph{recording} of a sampling outcome (i.e., which index $j$ was sampled) into a classical register. A typical example is the recording of a POVM $\{M_j\}_j$, which can be represented using Kraus operators $\sqrt{M_j}\otimes \ket{j}$. Channels of this form are referred to as ``fine-grained" quantum instruments\cite{chitambar2014everything}.

In addition to the recording of sampling outcomes into registers, our simulations use ``classical'' operations which act \emph{only} on classical registers. Unlike the classically conditioned quantum channels described above, these channels are able to change the register states. We model such operations as the probabilistic application of functions on the classical registers. To describe this, we consider a collection of functions $f_j$ acting on the space of register values. We wish to represent the probabilistic application of function $f_j$ with probability $p_j$. This can be modeled explicitly using Kraus operators
\begin{equation}
    \label{eq:classical-kraus}
    K_{j,l} = \sqrt{p_j}\sum_{\bar t\in \mathcal T_j^{(l)}} \ketbradt{f_j(\bar t)}{\bar t}
\end{equation}
where for each $j$, we have partitioned the input space of $f_j$ into disjoint subsets $\mathcal T_j^{(l)}$. The partition is chosen such that $f_j$ is \emph{injective} when restricted to a given subset of inputs $\mathcal T_j^{(l)}$. This guarantees that the resulting Kraus operators are trace preserving:
\begin{align}
    \label{eq:kraus-classical}
   \begin{split}
        \sum_{j,l} K_{j,l}^\dagger K_{j,l} & = \sum_j p_j \sum_{l}\sum_{\bar t, \bar t' \in \mathcal T_j^{(l)}} \ketbradt{\bar t'}{\bar t} \braket{f_j(\bar t')}{f_j(\bar t)}\\
        & = I
   \end{split}
\end{align}
where in the second line we used the injective property and the fact that $\sum p_j = 1$. As expected, the Kraus operators in~\eqref{eq:classical-kraus} can never generate superpositions of classical register states.

\subsection{Simulating incoherent systems \label{sec:simulating_incoherent}}

Here we formally define the notion of incoherence for both operators and quantum channels \cite{Q-coherence-review, quantifying_coherence, incoh-ops}. We then outline how the assumption of incoherence enables simulation of leakage without significantly increasing computational costs. We begin by considering a decomposition of
Hilbert space into a direct sum of subspaces,
\[
\mathcal{H}=\bigoplus_{n=1}^S\mathcal{H}_{n}.
\]
Let $P_{n}$ be the projection operator into subspace $\mathcal{H}_{n}$.
We define the \emph{dephasing channel} for the subspaces $\mathcal{H}_{n}$
as
\[
\Delta(\rho)=\sum_{n=1}^SP_{n}\rho P_{n}.
\]
This channel is equivalent to measuring which subspace a given state
is in and discarding that information; it destroys coherence between
states in different subspaces. For example, if we start with a pure
state
\[
\ket{\psi}=\alpha\ket{\psi_{a}}+\beta\ket{\psi_{b}}
\]
that is a coherent superposition of states from different subspaces
(${\ket{\psi_{a}}\in\mathcal{H}_{a}}$,${\ket{\psi_{b}}\in\mathcal{H}_{b}}$),
the action of $\Delta$ produces a mixture of these states:
\[
\Delta\l(\ketbrad{\psi}\r)=|\alpha|^{2}\ketbrad{\psi_{a}}+|\beta|^{2}\ketbrad{\psi_{b}}.
\]
Accordingly, we say a state $\rho$ is $incoherent$ with respect to
subspaces $\mathcal{H}_{n}$ if is invariant under the dephasing channel,
\[
\rho=\Delta(\rho)\,.
\]
A state $\rho$ is incoherent if and only if it is a statistical mixture
of states each contained in a (possibly different) single subspace $\mathcal{H}_{n}$. This is
equivalent to saying that it is block diagonal with respect to the
subspaces $\mathcal{P}_{n}$. More generally, we define \emph{any}
operator $A$ as incoherent with respect to subspaces $\mathcal{P}_{n}$
if $A=\Delta(A)=\sum_{n=1}^S P_{n}AP_{n}$.

We use the definition of incoherent states to describe the incoherent quantum channels. A quantum channel
$\mathcal{E}$ is incoherent \cite{quantifying_coherence} if it admits a set of Kraus operators $\{K_{j}\}_j$ such that, for all $K_j$ and any incoherent state $\rho=\Delta(\rho)$,
\begin{equation}
    \label{eq:kraus_dephasing}
    K_{j}\rho K_{j}^{\dagger}=\Delta(K_{j}\rho K_{j}^{\dagger})\,.
\end{equation} 
Such a set of Kraus operators has the property that, for any $K_j$ and subspace projector $P_n$,
 there exists another projector $P_m$ such that \cite{incoh-ops} 
\begin{equation}
    \label{eq:SSP}
    K_j P_n = P_m K_j P_n\,.
\end{equation}
In the context of a quantum trajectories simulation, this means that a state which starts incoherent will always remain incoherent, irrespective of which Kraus operator was sampled. We note that a given channel can have more than one distinct Kraus operator representation \cite{nielsen_chuang_2010}, and it is possible that not all representations of an incoherent channel satisfy Eq.~\eqref{eq:kraus_dephasing}.

The guarantee of incoherent operations can be used to improve both memory and time costs of quantum trajectories simulations. To demonstrate this it is convenient to introduce classical registers to the state vector to keep track of which subspace it is in. A general (possibly coherent) state may be decomposed as
\begin{equation}
    \ket{\tilde \psi}= \sum_{n=1}^S c_n\ket{\psi_n}\otimes \ket{n}
\end{equation}
where $\ket{\psi_n}\in \mathcal{H}_n$ are (normalized) states with support in a single subspace\citedfn{fn-standard-rep}. Each Kraus operator of an incoherent channel [with Kraus operators satisfying (\ref{eq:SSP})] can similarly be decomposed as
\begin{equation}
    \tilde K_j = \sum_{n=1}^S P_{m_j(n)}K_j P_n\otimes \ketbradt{m_j(n)}{n}
\end{equation}
where $m_j(n)$ is the unique index such that $P_m K_j P_n\neq 0$.
The action of a given Kraus operator on the state will then be
\begin{equation}
    \tilde K_j\ket{\tilde \psi} = \sum_{n=1}^S c_n P_{m_j(n)}K_j \ket{\psi_n}\otimes \ket{m_j(n)}.
\end{equation}
Notice that the classical register still reflects the supporting subspace. Importantly, the memory required to represent a vector $\ket{\psi_n}\otimes \ket{n}$ in a single subspace is 
$$
O\l( \text{dim} (\mathcal{H}_n) \r)+O(\log(S)),
$$
which can be exponentially smaller than ${\dim(\mathcal{H})=\sum_{n}\dim \mathcal{H}_n}$. The time required to apply the operation $P_{m_j(n)}K_j \ket{\psi_n}$ can be similarly suppressed, since only the input and output subspaces need to be represented. Thus if the initial quantum state has support in a single subspace (i.e. $c_n\neq 0$ for exactly one $n$), the cost of a trajectories simulation will be set by $\text{max}\l(\text{dim}( \mathcal{H}_n)\r)$ instead of $\text{dim}(\mathcal{H})$.

Now let us consider the case of leakage outside of the computational (qubit) subspace. In the simplest case we model a 3-level qudit, with Hilbert space decomposed into a `computational' and `leakage' subspace
\begin{equation*}
\begin{split}
 & \mathcal{H} = \mathcal{C} \oplus \mathcal{L} \\
& \mathcal{C} =  \text{span}\{\ket{0},\ket{1}\} \\
& \mathcal{L} =  \text{span}\{\ket{2}\}\,.   
\end{split}
\end{equation*}
Here the states $\ket{0},\ket{1},\ket{2}$ represent the lowest energy eigenstates of the transmon Hamiltonian. We say that leakage in our system is incoherent if the system state only ever has support in $\mathcal{C}$ \textit{or} $\mathcal{L}$.
Using the construction above, this means the memory requirements for such a simulation are still $O(2^m)$ for $m$ such qudits instead of $3^m$. 

\subsection{Subspace Twirling Approximation}
\label{sec:STA}

In cases where a quantum channel is \emph{not incoherent} with respect to the relevant subspaces, we describe an approximation to generate an equivalent incoherent channel. We will use this approximation in the context of leakage in superconducting qubits, as indeed some leakage mechanisms can be coherent in nature.
The approximation is motivated by the fact that, due to the transmon nonlinearity, the leaked states rapidly accumulate a phase (relative to the computational basis states) over the course of a single round of error correction. For state $\ket{2}$ this phase is of the form $\exp(-i 2 \pi \eta t_{r})$, where $t_r\sim 1 \mu$s is the duration of an error correction round. For typical nonlinearities $\sim 200 $MHz the phase oscillates rapidly, which suggests that we can treat it as random. To model this random accumulation of phase between subspaces, we consider a phase unitary
\[
U(\bar{\phi})=\sum_{n}e^{i\phi_{n}}P_{n}
\]
which assigns a distinct phase $\phi_{n}$ to each subspace. The
Subspace Twirling Approximation (STA) is equivalent twirling over the set of unitaries $U(\bar{\phi})$ \cite{dankert2009exact,bartlett2007reference}.
This means conjugating with respect to  $U(\bar{\phi})$ and its inverse and averaging over independent and uniformly distributed phases
$\phi_{n}$,
\begin{equation}
\label{eq:STADef}
\mathcal{E}\rightarrow\mathcal{E}_{STA}=\mean{\hat{U}_{-\bar{\phi}}\circ\mathcal{E}\circ\hat{U}_{\bar{\phi}}}_{\bar \phi},
\end{equation}
where $\hat{U}_{\bar{\phi}}$ is the quantum channel associated with the unitary $U(\bar \phi)$.

Expanding this out, we have
\begin{align*}
\mathcal{E}_{STA}(\rho) & =\sum_{j}\mean{U(-\bar{\phi})K_{j}U(\bar{\phi})\,\rho\,U^\dag (\bar{\phi})K_{j}^{\dagger}U^{\dagger}(-\bar{\phi})}_{\bar{\phi}}\\
 & =\sum_{m,m',n,n'}\sum_{j}\mean{e^{i(\phi_{n}-\phi_{n'}-\phi_{m}+\phi_{m'})}}_{\bar{\phi}}\\
 & \times P_{m}K_{j}P_{n}\,\rho\,P_{n'}K_{j}^{\dagger}P_{m'}.
\end{align*}
Observe that the average $\mean{e^{i(\phi_{n}-\phi_{n'}-\phi_{m}+\phi_{m'})}}_{\bar{\phi}}$
is exactly zero unless the indices satisfy 
\[
\phi_{n}+\phi_{m'}=\phi_{m}+\phi_{n'}
\]
i.e. unless I) $n=n',m=m'$ or II) $m=n,m'=n'$ . We can break up the sum into these cases
\begin{align*}
\mathcal{E}_{STA}(\rho)= & \sum_{m,n,m\neq n}\sum_{j}P_{m}K_{j}P_{n}\,\rho\,P_{n}K_{j}^{\dagger}P_{m}\\
 & +\sum_{j}\l(\sum_{n}P_{n}K_{j}P_{n}\r)\,\rho\,\l(\sum_{n'}P_{n'}K_{j}^{\dagger}P_{n'}\r)
\end{align*}
where the second term corresponds to II and the first to I and not II. The Kraus operators generated under the STA can take two possible forms. First there
are cross terms
\begin{equation}
P_{m}K_{j}P_{n}\label{eq:KrausCrossSTA}
\end{equation}
which can be associated with incoherent transitions. The second form
is just the block diagonal part of $K_{j}$,
\begin{equation}
\Delta(K_j) = \sum_{n}P_{n}K_{j}P_{n}\label{eq:KrausDiagSTA}
\end{equation}
which preserves each subspace. 

We note some important properties of the STA. First, although the Kraus
operators generated in Eqs.~(\ref{eq:KrausCrossSTA}) and (\ref{eq:KrausDiagSTA})
depend on the specific choice of Kraus operators $K_{j}$ used to
represent $\mathcal{E}$, the resulting channel in (\ref{eq:STADef})
is \emph{representation independent}. Thus, if $\{K_{j}\},\{K_{j}'\}$ are
two sets of distinct Kraus representations of $\mathcal{E}$, the resulting
Kraus operators from equations (\ref{eq:KrausCrossSTA}), (\ref{eq:KrausDiagSTA})
will correspond to the \emph{same channel}. 
Second, observe that both forms of Kraus operators for $\mathcal{E}_{STA}$
satisfy Eq.~\eqref{eq:SSP}.
Therefore $\mathcal{E}_{STA}$ is an incoherent channel for every
$\mathcal{E}$. In fact, $\mathcal{E_{STA}}$ satisfies an even stronger
condition: Inserting the Kraus operators of $\mathcal{E}_{STA}$ into~\eqref{eq:SSP}, we see that for each fixed $j$, the function $m(j,n)$
is \emph{injective} in $n$ (i.e., each output has at most one unique
input). This classifies $\mathcal{E}_{STA}$ within the \emph{strictly incoherent
operations} \cite{incoh-ops}, which is a strict subset of the incoherent operations \cite{Q-coherence-review}.
Third, we note that if one of the projectors $P_n=P_C$ represents the full computational subspace and $\mathcal E$ denotes an error channel, then the STA \emph{preserves} the process fidelity of that channel 
\begin{equation}
    F_{\text{pro}}(\mathcal E_{STA},\mathcal I) = F_{\text{pro}}(\mathcal E,\mathcal I)\,.
\end{equation}
Here we are using the generalized definition of process fidelity \cite{wood2018quantification}, which for Kraus operators $K_j$ can be computed as \cite{pedersen2007fidelity}
$$F_{\text{pro}}(\mathcal E,\mathcal I) \equiv \frac{1}{D^2}\sum_j \l|\mbox{Tr} [P_C K_j]\r|^2\,,$$
where $D=2^k$ is the size of the (local) computational subspace on which the channel acts. Additionally, we point out that the STA is a weaker approximation than simply dephasing the system,
\begin{equation*}
    \mathcal E\rightarrow \Delta \circ \mathcal E\,,    
\end{equation*}
since STA preserves the identity channel.
Finally, we comment on the connection of the STA to twirling\cite{bartlett2007reference}. The unitary channel $\hat U_{\bar \phi}$ is a representation of $U(1)^{\times m}$, where $m$ is the number of distinct subspaces $\mathcal H_n$.  The twirling operation represented by equation~\eqref{eq:STADef} ensures that $\mathcal {E}_{STA}$ is a morphism of this representation, which means that it commutes with $\hat U_{\bar \phi}$ for every phase $\bar \phi$. This guarantees that $\mathcal {E}_{STA}$ preserves incoherence of states: observe that a state is incoherent with respect to the subspaces $\mathcal H_n$ if and only if it is invariant under $\hat U_{\bar \phi}$, i.e.
\[
\hat U_{\bar \phi} (\rho) = \rho
\]
for all $\bar \phi$. Because $\mathcal {E}_{STA}$ commutes with $\hat U_{\bar \phi}$, we have that
$$ \mathcal {E}_{STA} (\rho) =  (\mathcal {E}_{STA} \circ  \hat U_{\bar \phi}) (\rho) =( \hat U_{\bar \phi} \circ \mathcal {E}_{STA})\, (\rho),$$ 
and is therefore still an incoherent state.

\section{Incoherent leakage in error correcting codes \label{sec:STA-justification}}

The properties of typical superconducting qubit logical memory experiments \cite{state-preservation-rc-google, scaling_SC_Google_2022, marques2022logical, zhao2022realization} provide several physical justifications for the Subspace Twirling Approximation. First, in many cases, the physical mechanism for leakage is already an approximately incoherent process. This is typically the case when leakage is caused by decay, dephasing, or heating processes. Such channels act under qubit idling \cite{chen_measuring_2016} or during the CZ gate \cite{tunable-coupler-Yan, tunable-coupler-cphase, foxen_demonstrating_2020}. Though the STA on such channels is actually non-trivial, in practice it is a very accurate approximation in the context of error correction, as we will see below.
Conversely, a unitary leakage source (for example a control error), is generally not be well approximated by the STA. We refer to such channels as sources of `coherent leakage'.
In most gate implementations, coherent leakage errors are actually quite small, as they are minimized through calibration~\cite{foxen_demonstrating_2020}. Yet even in this case, the repeated rounds of stabilizer measurement can be used to detect when leakage in a qubit has occurred~\cite{leakage-detection}. Accordingly, as demonstrated below, the coherence between leakage and computational subspaces decays exponentially in the number of stabilizer measurements. This suggests that we can model the coherence of leakage as a short-lived process:  although it has an impact on leakage population growth over the course of a single round of error correction, over the course of multiple rounds it can be treated as incoherent. This motivates the idea that, in the context of error correction experiments, we can model a coherent leakage mechanism using an effective thermal model (which is amenable to the STA).

\subsection{Decay of leakage superposition under stabilizer measurement \label{sec:coherence-decay}}

In order to motivate the validity of the STA within typical error correction simulations, we show that repeated
stabilizer measurements have the effect of decohering the leakage
and computational subspaces. 
This is analogous to the observation that for coherent (and non-Pauli) errors acting on the computational subspace, stabilizer measurements effectively twirl the noise into a Pauli error model \cite{tomita_low-distance_2014,katabarwa_logical_2015,cai_constructing_2019}.
To focus on the dominant effects, we assume that no leakage is caused by the CZ unitary. Furthermore we ignore any leakage in the measure qubit\citedfn{fn-dq-leak}, so that the
unitary can be simply expressed as 
\[
U_{CZ}=\ketbrad{0}_{d}\otimes\mathbb{I}_{m}+\ketbrad{1}_{d}\otimes Z_{m}+\ketbrad{2}_{d}\otimes\exp(i\phi Z_{m}/2)
\]
where the first coordinate corresponds to the (possibly leaky) data
qubit and the second coordinate corresponds to the measure qubit.
As seen in the above equation, we assume that if the data qubit is state
$\ket{2}$, the Hamiltonian dynamics of the CZ gate causes the
measure qubit to acquire a phase $\phi$ between the $\ket{0}$
and $\ket{1}$ states. The stabilizer measurement procedure can then be expressed as a product of unitaries
\[
U_{S}=H_{m}\l(U_{CZ}\otimes \mathbb{I}_r\r)U_{rest}H_{m}
\]
where $H_{m}$ is a Hadamard on the measure qubit and 
\[
U_{rest}=\mathbb{I}_{d}\otimes(\mathbb{I}_{m}\otimes P_{0,r}+Z_{m}\otimes P_{1,r})
\]
represents the other CZ unitaries applied to the other data qubits
of the stabilizer (denoted by subscript $r$). The projection operators $P_{0,r}$ and $P_{1,r}$
represent the $-1$ and $+1$ eigenvalue subspaces of the rest of
the stabilizer (for simplicity we assume the other data qubits do
not leak). Conjugating the Hadamards into the other unitaries gives
\begin{align*}
U_{S}= & (\ketbrad{0}_{d}\otimes\mathbb{I}_{m}+\ketbrad{1}_{d}\otimes X_{m}+\ketbrad{2}_{d}\otimes\exp(i\phi X_{m}/2))\\
 & \times(\mathbb{I}_{m}\otimes P_{0,r}+X_{m}\otimes P_{1,r})
\end{align*}
where we have omitted the trivial identity tensor products. 

We can now extract the Kraus operators acting on the data qubits by
assuming that the measure qubit is prepared in the $\ket{0}$ state
and read out in the computational basis. This gives
\begin{align*}
K_{0} & =\bra{0}_{m}U_{S}\ket{0}_{m}=L_{0,d}\otimes P_{0,r}+L_{1,d}\otimes P_{1,r}\\
K_{1} & =\bra{1}_{m}U_{S}\ket{0}_{m}=L_{0,d}\otimes P_{1,r}+L_{1,d}\otimes P_{0,r}
\end{align*}
where
\begin{align*}
L_{0,d} & =\ketbrad{0}_{d}+\cos(\phi/2)\ketbrad{2}_{d}\\
L_{1,d} & =\ketbrad{1}_{d}+i\sin(\phi/2)\ketbrad{2}_{d}.
\end{align*}
As expected, $K_{0}$ and $K_{1}$ reflect the parity between the
data qubit $Z$ eigenvalue (excluding leakage) and the rest of the stabilizer. 

The Kraus operators can help explain the decay of coherence between
leakage and computational subspaces. Suppose, for example, that stabilizer
value $j$ was measured $m$ times in succession. This would correspond to the Kraus operator
\begin{equation}
    \label{eq:Kjm}
    K_{j}^{m}=L_{0,d}^{m}\otimes P_{j,r}+L_{1,d}^{m}\otimes P_{1-j,r}
\end{equation}
For angles $\phi$ not equal to multiples of $\pi$, observe that
\begin{align*}
L_{0,d}^{m} & =\ketbrad{0}_{d}+\cos(\phi/2)^{m}\ketbrad{2}_{d}\stackrel{m\rightarrow\infty}{\rightarrow}\ketbrad{0}_{d}\\
L_{1,d}^{m} & =\ketbrad{1}_{d}+(i\sin(\phi/2))^{m}\ketbrad{2}_{d}\stackrel{m\rightarrow\infty}{\rightarrow}\ketbrad{1}_{d}
\end{align*}
hence 
\[
K_{j}^{m}\stackrel{m\rightarrow\infty}{\rightarrow}\ketbrad{0}_{d}\otimes P_{j,r}+\ketbrad{1}_{d}\otimes P_{1-j,r}\,.
\]
The operator in this limit vanishes on state $\ket{2}$. Thus, ignoring any other errors in
the system, a sequence of many repeated stabilizer measurements indicate
that the data qubit is in the computational subspace with high probability.
Similarly, suppose that stabilizer measurement $0$ was followed by
stabilizer measurement $1$. This would correspond to the Kraus operator
\begin{align}
\label{eq:K1K0}
\begin{split}
    K_{1}K_{0} & =L_{1,d}L_{0,d}\otimes P_{0,r}+L_{0,d}L_{1,d}\otimes P_{1,r}\\
 & =L_{1,d}L_{0,d}\\
 & =\frac{i}{2}\sin(\phi)\ketbrad{2}_{d}
\end{split}
\end{align}
where we have used the fact that $L_{0,d}^{m}$ and $L_{1,d}^{m}$
commute. Therefore, ignoring any other errors in the system, any change
in stabilizer measurements indicates that the data qubit has leaked. 

\begin{figure}
    \centering
    \includegraphics[width=0.98\columnwidth]{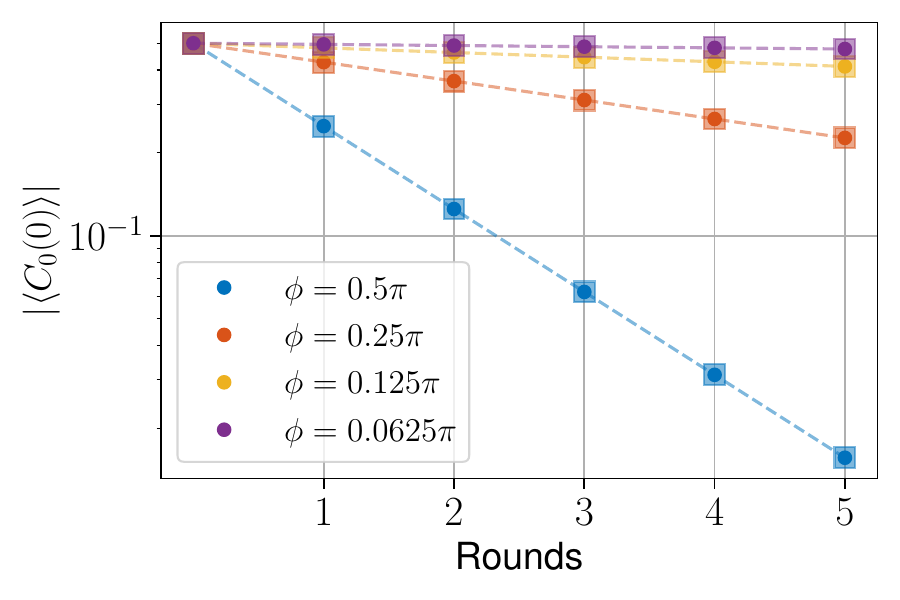}
    \caption{\textbf{Repetition code coherence decay under stabilizer measurements.} For a distance 3 repetition code, where the central data qubit is initially in state $(\ket{0} + \ket{2})/\sqrt{2}$ (and the other data qubits in $\ket{0}$), we plot the decay in the qubit-leakage coherence of the central data qubit, as a function of the number of rounds, where each round consists of 2 CZ gates on the qubit. The circles are the data from simulation, and the squares with dash lines the value from theory, Eq.~\eqref{eq:coherence_decay}.}
    \label{fig:rc_coherence_decay}
\end{figure}

This analysis can be used to predict the decay of coherence between leakage and computational subspaces. In the absence of any other noise in the system, repeated stabilizer measurements over $m$ rounds correspond to products of Kraus operators 
$$K_{\bar j}= K_{j_m} K_{j_{m-1}}\,...\,K_{j_1},\,$$
where $j_k\in\{0,1\}$. Now consider the observables 
$$C_c(\theta)=e^{i \theta} \ketbradt{c}{2}_d \otimes \mathbb{I}_r + \mbox{h.c.}$$
for $c=0,1$, which serve as measures of the coherence between the leakage and computational subspace of the data qubit. In the Heisenberg picture, after $m$ rounds of stabilizer measurement these observables become 
$$\sum_{\bar j} K_{\bar j}^\dagger C_c(\theta) K_{\bar j}\,.$$ 
From Eqs.~\eqref{eq:Kjm} and \eqref{eq:K1K0} we see that the summand vanishes unless $\bar j = (0,0,\,...\,0)$
or $\bar j = (1,1,\,...\,1).$
Substituting Eq.~\eqref{eq:Kjm} for these values of $\bar j$, one may show that
\begin{equation}
\begin{split}
C_c(\theta) & \rightarrow \sum_{j=0,1} \l(K_j^\dagger\r)^m C_c(\theta) K_j^m\\
& = r_c(m) C_c(\theta + \delta_{c,1} m\pi/2)
\end{split}
\label{eq:coherence_decay}
\end{equation}
where 
\begin{equation*}
    r_c(m) =\l\{ \begin{tabular}{cc}
     $\cos(\phi/2)^{m}$,& $c=0$  \\
     $\sin(\phi/2)^{m}$,& $c=1.$ 
\end{tabular}\r.
\end{equation*}
As expected, for any $\phi$ not an integer multiple of $\pi$, all coherence measures $C_c(\theta)$ decay exponentially in the number of measurements.  (If the data qubit interacts with $k$ distinct measure qubits per round, $m=kr$ for $r$ rounds.) We see the case $\phi=\pi/2$ results in a uniform decay of $2^{-m/2}$. 
We show numerical and theoretical results for the repetition code in Fig.~\ref{fig:rc_coherence_decay}. In practice, even when $\phi$ is an integer multiple of $\pi$, it is straightforward to modify the error correction circuit to ensure that every coherence $C_c(\theta)$ operator decays. For example, this can be done by injecting a logical $X$ operator between error correction rounds \cite{scaling_SC_Google_2022}, which flips all data qubits ($c\leftrightarrow 1-c$) in the standard basis.

We can use our analysis to describe the patterns of measurement outcomes given a leaked data qubit. Suppose the data qubit starts in the state $\ket{2}$ and the rest of the stabilizers are in the -1 eigenspace.
We can compute the probability that we observe exactly $m_0$ measurements of 0 out of $m$ stabilizer measurements. From the above, this is simply
\begin{equation*}
 p_{m_0}^{(-1)} = {m \choose m_0} \cos(\phi/2)^{2m_0}\sin(\phi/2)^{2(m-m_0)}.
\end{equation*}
Similarly, if the rest of the stabilizers were in the $+1$ eigenspace, the exponents should be switched.
This could be used to infer the occurrence of a leakage event from the measurement statistics. Moreover, since $\phi$ dictates the probability of stabilizer outcomes under leakage, which are totally random for $\phi=\pi/2$, it may be possible to reduce the logical error caused by leakage, by decreasing $\phi$ (see Fig.~\ref{fig:dist_5_comparison}). 

\subsection{Numerical validation of the STA \label{sec:STA_validation}}
To test the validity - and limitations - of the STA, we perform simulations corresponding to two distinct models of leakage, for local dimension $D=3$. Our main results are for the distance 3 rotated surface code, though in App.~\ref{sec:rep-code-STA} we also present results for the repetition code up to distance 9.

The first model considered assumes leakage is caused by thermal excitations \cite{chen_measuring_2016}. Such processes can be modeled using a Markovian master equation with Lindblad operators of the form
\begin{equation}
    \label{eq:heating-leakage}
    L_h = \frac{1}{\sqrt{T_h}}a^\dag\,,
\end{equation}where $T_h$ is the characteristic heating time scale. We incorporate the heating Lindblad operators into our overall decoherence model which evolve over each gate (described below). Note that such a model can result in computational-leakage coherence, and so the STA acts non-trivially on the channel. However, for long enough time-scales the STA approaches the exact dynamics (since the steady state is diagonal).

The second model we consider assumes that leakage is the result of purely coherent transitions during the CZ gate, of the form
\begin{equation}
    \label{eq:cz:coherent-leakage}
    |11\rangle \rightarrow \sqrt{1-p}\ket{11} + e^{i\phi_{11,02}}\sqrt{p}\ket{02},
\end{equation}
for parameters $p,\phi_{11,02}$ discussed below.
As demonstrated in e.g., Refs.~\cite{scaling_SC_Google_2022, foxen_demonstrating_2020}, this transition occurs during the diabatic CZ gate, and, as we will show, can, in principle, generate significant coherence between computational and leakage subspaces. (In the simulations in this section this is partly because we pick larger than typical values of the coherent leakage rate $p$, for demonstrative purposes.)
In these simulations, we limit leakage to the second level $\ket{2}$.

In both cases considered we also include the effect of decoherence. This is incorporated as $T_1$-type decay (Linblad operators proportional to the annihilation operator, $L_1 = {a}/{\sqrt{T_1}}$) and $T_\phi$ white noise dephasing (Linblad operators proportional to the number operator, $L_\phi = {n}\sqrt{2/{T_\phi}}$). To model these effects in the discrete time quantum trajectories, after each unitary gate or idle of duration $t$ we apply an additional noise channel:
\begin{equation}
\label{eq:lindbladian}  
    e^{t ( \mathcal D[L_1] + \mathcal D[L_\phi])}\,,
\end{equation}
where $\mathcal D[L](\rho) = L \rho L^\dagger - \frac{1}{2}\{L^\dagger L, \rho\}$ is the standard Lindblad form superoperator. (The heating model also includes the Lindblad operator $\mathcal D[L_h]$.) We pick parameters in line with recent experiments~ \cite{state-preservation-rc-google, scaling_SC_Google_2022, google-leakage-2016}. In particular, $T_1=20\mu$s (determining $1\rightarrow 0$ transition), and $T_\phi=80\mu$s. To avoid leakage accumulation, we also set the time-scale for the $2\rightarrow 1$ transition to be $T_L=10\mu$s. For reference, each round of the surface code has duration $1.1 \mu s$ when using realistic gate durations.
See App.~\ref{sec:simulation} for more information and implementation details, and App.~\ref{sec:error-rate} for information on fitting the error rate.
The dominant contribution of the decoherence model occurs during measurement/reset of the auxiliary  qubits, as the time of such operations is much longer than regular gate times (e.g., 900 ns compared to 25 ns).

\subsubsection{Thermal model STA simulations}
\label{sec:thermal-STA}
Here we validate the STA in a setting where leakage arises from external heating modelled through a Lindblad master equation, with a Lindblad operator $a^\dag/\sqrt{T_h}$. In these simulations we set $T_h=1000\mu$s, resulting in characteristic $1\rightarrow 2$ transition time of $500\mu$s (cf. Ref.~\cite{scaling_SC_Google_2022} which has a time of $700\mu$s), and a steady state leakage population close to $1\%$.

Note that our model of thermal heating, described by the Lindblad operator $\propto a^\dagger$, does not take into account the qubit non-linearity. This means that we are likely \emph{overestimating} any coherent effects associated with such leakage. Properly accounting for the non-linearity will cause coherent transitions $\ketbra{2}{1}\rho\ketbra{0}{1}$ in the Lindblad master equation to oscillate at a frequency $2\pi \eta$, much faster than the heating rate $1/T_h$. Such transitions could formally be disregarded under a rotating wave approximation. Nevertheless, even without taking the non-linearity into account, we find that the STA accurately models error correction properties of interest.

\begin{figure*}
    \centering
    \includegraphics[width=1.75\columnwidth]{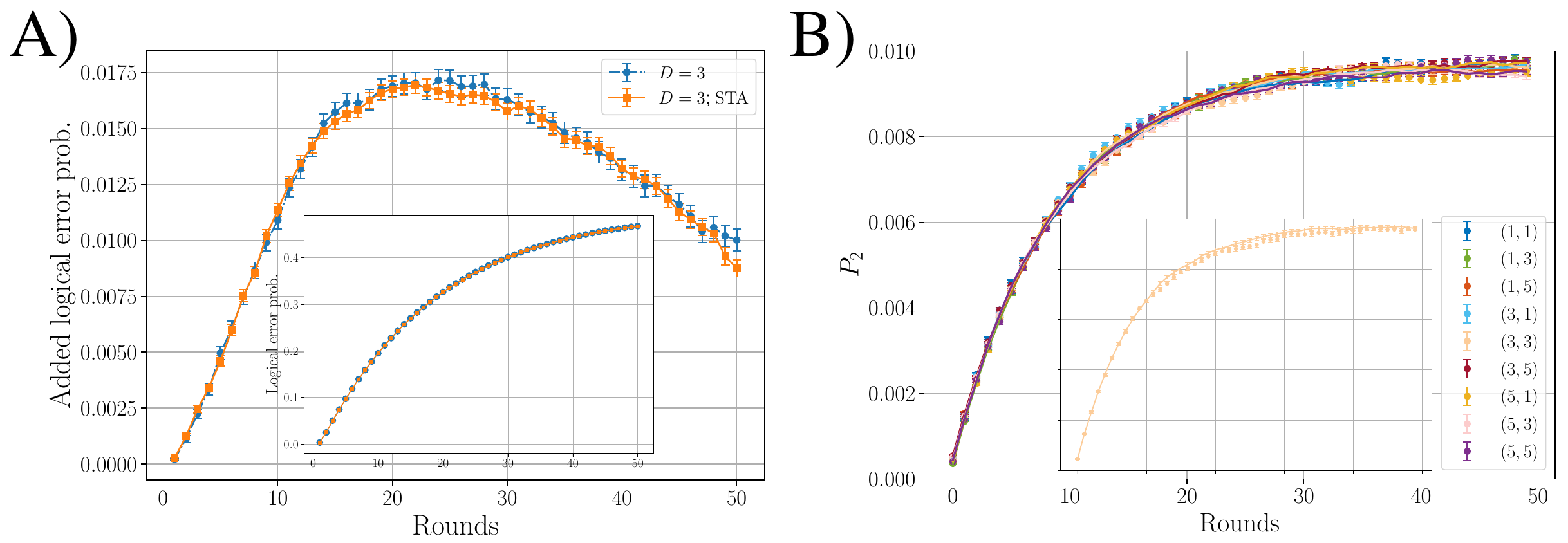}
    \caption{\textbf{STA for thermal model in distance-3 surface code.} A) (Main figure) The logical error probability added by leakage, for the exact (three-level) simulation, and the same under the STA. We subtract the total logical error probability from that observed in the zero-leakage limit ($T_h \rightarrow \infty$ in Eq.~\eqref{eq:heating-leakage}). The LER added by leakage are fitted as  $0.266\pm 0.009\%$ and $0.275 \pm 0.012\%$ with and without the STA respectively. Error bars are the standard error over at least $10^6$ samples. (Inset) The total logical error probability.  B)  The leakage population for the data qubits (labeled according to Fig.~\ref{fig:SC_schematic}). The circles are the data points without the STA, and the lines are the data with the STA applied. We see the $|2\rangle$ state population saturates close to $T_L/T_h =1\%$ for each qubit [see Eq.~\eqref{eq:P2_inf}]. Inset shows data for the central $(3,3)$ data qubit (see Fig.~\ref{fig:SC_schematic}).}
    \label{fig:thermal-logical-err}
\end{figure*}

In Fig.~\ref{fig:thermal-logical-err} A), we study the contribution of leakage to the logical error probability. The STA and full simulation results agree to high precision. We particularly focus on the \emph{added} logical error rate (LER) due to leakage (App.~\ref{sec:error-rate}). This is the excess LER when comparing equivalent simulations with and without leakage. We observe leakage added logical error rates of $0.275 \pm 0.012\%$ and  $0.266\pm 0.009 \%$, without and with the STA respectively, agreeing within the expected error range (for reference, the leakage-free LER is $2.47\pm 0.010 \%$). Fig.~\ref{fig:thermal-logical-err} B) shows leakage populations $P_2$ for the two models, again demonstrating fairly high agreement; the relative error is within around 5\%, and the fluctuations are likely caused by finite sampling (here we use at least $10^6$ trajectories). In App.~\ref{sec:STA-additional} we additionally verify that detection event statistics agree between the models.

\subsubsection{Coherent leakage STA simulations \label{sec:STA-validation-coherent}}
In this section, we model all leakage as a strictly coherent (unitary) process occurring during the CZ gate, which therefore necessarily generates coherence between the qubit and leaked subspace during the simulation. The decoherence model of Eq.~\eqref{eq:lindbladian} does \emph{not} in include a heating contribution. This is a nonphysical assumption \cite{chen_measuring_2016} with the aim to demonstrate the limitations of the STA applied in the limit of strong coherent leakage. Indeed, we will pick the CZ leakage rate to be larger than found in recent experiments \cite{scaling_SC_Google_2022}. The results here are as a result strongly dependent on the non-linearity, which we set to $\eta=0.3$GHz here, though provide additional results in App.~\ref{sec:STA-additional}.

As mentioned above, our coherent leakage model results in transitions of the form $|11\rangle \rightarrow \sqrt{1-p}\ket{11} + e^{i\phi_{11,02}}\sqrt{p}\ket{02}$. We assume only one of the two qubits coherently leaks during the CZ gate, as would be expected from a miscalibration of the gate timings. (This is true in the case of ``diabatic'' gates~\cite{foxen_demonstrating_2020}.)
Since the auxiliary (measure) qubits are reset to state $\ket{0}$ each round, we set up the transitions to occur \emph{only on the data qubits}. This is to give the worst-case scenario for leakage in the surface code: fully coherent leakage that is not removed by external mechanisms and can persist for multiple rounds.
We show in App.~\ref{sec:STA-additional} that the `transition phase' $\phi_{11,02}$ has no obvious effect on the results, and so we set $\phi_{11,02}=0$ here. We pick the transition probability $p=2.4\times 10^{-3}$, noticeably larger than found in current experiments for CZ heating (cf. Ref.~\cite{scaling_SC_Google_2022} with a value $p=8\times 10^{-4}$), as such the coherent effects will be more pronounced.

\begin{figure*}
    \centering
    \includegraphics[width=2\columnwidth]{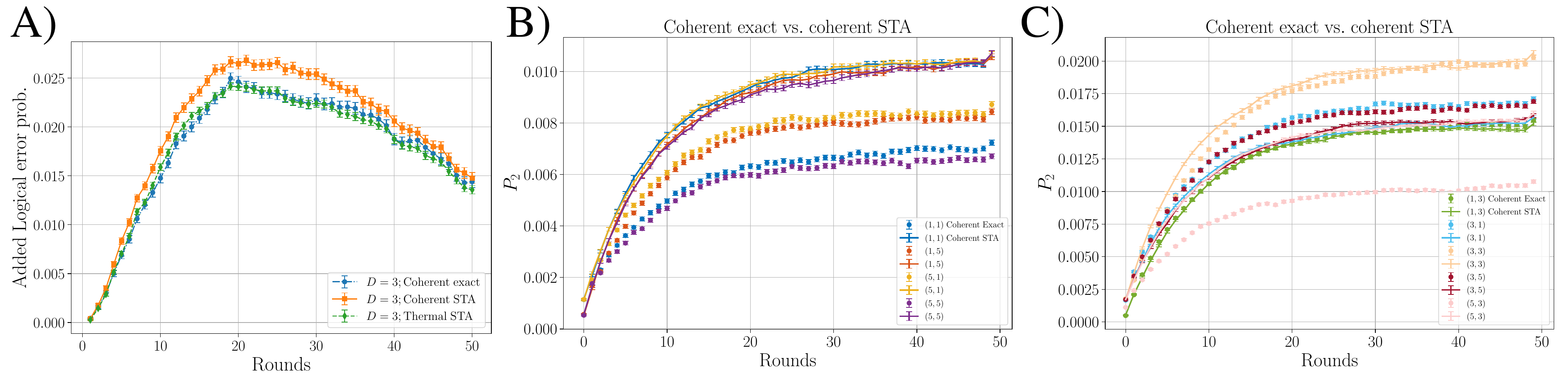}
    \caption{\textbf{STA for coherent CZ leakage in distance-3 surface code.} A) The logical error probability added by leakage for the coherent leakage model for the exact simulation (blue/circles), the case with STA (orange/squares), and where we make the `thermal approximation' as described in the main text (green/diamonds).  We subtract the total logical error probability from that observed in the zero-leakage limit ($p=0$ in Eq.~\eqref{eq:cz:coherent-leakage})
     The extracted leakage added logical error rates are $0.384 \pm 0.015\%$, $0.404\pm 0.014\%$, and $0.365\pm 0.010\%$, respectively. Error bars are the standard error over at least $7\times10^5$ samples.  B) The leakage population for the four corner data qubits (labeled according to Fig.~\ref{fig:SC_schematic}). The circles are the data points without the STA, and the lines are the data with the STA applied. C) The same as B), but for the central [(3,3)] and edge data qubits.}
    \label{fig:cz-logical-err}
\end{figure*}

In Fig.~\ref{fig:cz-logical-err} we compare the performance of the surface code between the coherent leakage model and its STA. In contrast to the thermal model, we see clear and systematic discrepancies. This is most obvious in the leakage populations over time; in Figs.~\ref{fig:cz-logical-err} B, C, certain data qubits obtain significantly higher or lower steady state populations under the STA [e.g. the qubit (5,5)]. Note that we have grouped data in the figures B and C for better readability; figure B corresponds to the four corner data qubits (which only undergo two CZ gates per round), and figure C the remainder of the data qubits, which see three or four CZ per round (see Fig.~\ref{fig:SC_schematic}). In addition to the general observation that qubits with a greater number of CZ gates typically obtain higher leakage, we can also see a decidedly coherent effect, that qubits with the same number of CZ gates may have distinct leakage populations [e.g.~qubits (5,3) and (3,1)]. In contrast, the STA data mostly depends only on the number of CZ gates.

The disagreement between the STA (orange/square) and exact  (blue/circle) simulations in the coherent leakage model is less pronounced when considering the logical error probability, Fig.~\ref{fig:cz-logical-err}A. Even so, the STA leads to a systematic overestimation of the logical error probability in this case, with added logical error rates 0.404$\pm$0.014\% and $0.384\pm 0.0015\%$) with and without the STA, respectively (n.b.~the leakage free logical error rate is $2.36\pm 0.012\%$). In App.~\ref{sec:STA-additional} we demonstrate that this discrepancy is a genuine coherent effect. In particular, Fig.~\ref{fig:eta_LEP} shows the sensitivity of the LER to the non-linearity $\eta$ in the exact simulation. In contrast, the STA is agnostic to $\eta$ (by definition). We further show in Fig.~\ref{fig:leakage-eta} that the qubit leakage populations strongly depend on the non-linearity (in this section, we use $\eta=300$MHz). This arises since the relative phase accumulated between state $\ket{2}$ and the computational subspace during a fixed time interval is proportional to the non-linearity $\eta$. Under the STA, this phase has no physical effect because the qubits are never in superposition of leaked and computational state.

\begin{figure*}
    \centering
    \includegraphics[width=1.5\columnwidth]{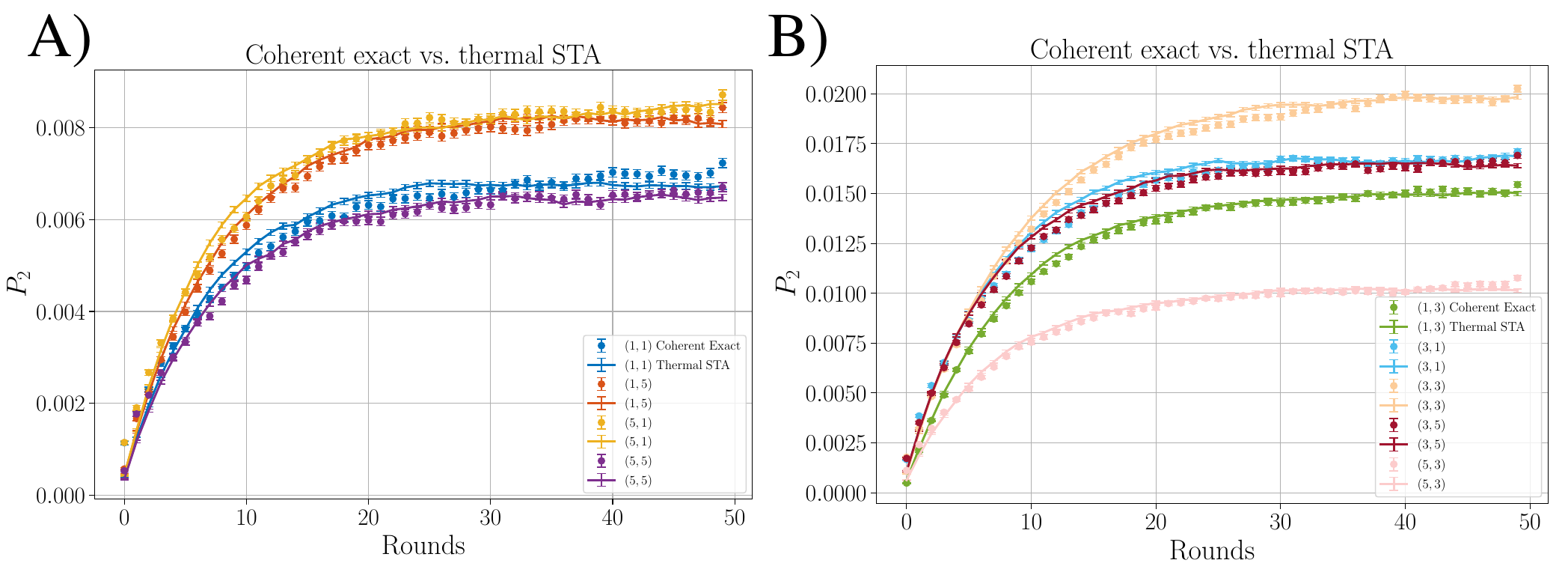}
    \caption{\textbf{STA for thermal approximation in distance-3 surface code.} 
    A) Leakage population for the corner data qubits, and B) for the remaining data qubits. Circular data points are the `exact' results (from the full three-level simulation) and the lines are after performing the thermal approximation (with STA) as described in the main text. Data collected over at least $7\times 10^5$ samples.}
    \label{fig:cz-map-logical-err}
\end{figure*}

While in the limit of \emph{purely coherent} (unitary) leakage, the STA has an effect on surface code logical error and leakage populations, it is possible to construct an effective thermal model which better agrees with the exact case, and therefore permits for accurate application of the STA. 
We motivate this model as follows. Although the STA is not accurate at the level of a single (coherent) operation, the effect of coherent leakage over multiple rounds of error correction can still be modeled using incoherent operations. This is because the stabilizer measurements themselves cause fast dephasing between the leakage and computational subspaces (Sect.~\ref{sec:coherence-decay}), so that significant coherence is never allowed to build up. As coherence between the computational and leakage subspaces is suppressed, we expect leakage populations to grow in an approximately incoherent/classical fashion.
To take advantage of this observation, we construct a thermal model of leakage - where STA is known to work well - whose parameters are fitted to match the observed leakage population in the coherent leakage model. In particular, given the qubit leakage populations over time, we extract an effective classical $1 \rightarrow 2$ transition rate that parameterizes a  thermal model (i.e.~with Lindblad heating) as discussed in  App.~\ref{sec:classical-rate}. The results are shown in Fig.~\ref{fig:cz-logical-err}A and Fig.~\ref{fig:cz-map-logical-err}. We see the green/diamonds of the `thermal approximation' in the figure (\ref{fig:cz-logical-err}A) much more accurately match the blue/circles (exact simulation), outperforming the naive application of the STA (orange/squares). 
Although the LER added by leakage does not show much improvement under the thermal approximation ($0.365\pm 0.010\%$ vs.~the exact $0.384\pm0.015\%$, and $0.404\pm0.014$ for STA), this apparent discrepancy is consistent within the error bars.
To quantify the improvement in the leakage-added logical error probability we compute the percentage error (averaged over all rounds) to the exact simulation for the data in Fig.~\ref{fig:cz-logical-err}A, which gives $2.9\%$ and $11.0\%$ for the thermal approximation and the `naive' STA respectively.
Moreover, in Fig.~\ref{fig:cz-map-logical-err} we see the individual leakage probabilities show a significant improvement in accuracy compared to those in Fig.~\ref{fig:cz-logical-err}. In particular, the thermal approximation leakage populations track very closely for all qubits, in stark contrast to the naive application of the STA.

Our simulations suggest that, even in the case where the STA is not directly applicable -- due to coherent effects which are by definition not captured by the STA -- an effective model of `thermal leakage' can be constructed for which the STA performs reasonably well. Specifying the model  requires knowledge of the leakage populations for the data qubits, as outlined in App.~\ref{sec:classical-rate}. Since this information is accessible experimentally, this approximation may allow for more efficient quantum simulations, despite the presence of coherent leakage. Furthermore, whilst it is clearly possible to devise nefarious (coherent) models of leakage such that the STA is inaccurate, we find that for physically realistic models of leakage in superconducting qubits, which includes both coherent and thermal contributions, using the STA \emph{without} the effective model still provides accurate predictions. We provide this data in App.~\ref{sec:STA-physical} for repetition codes up to distance 9, and the surface code at distance 3.

\begin{figure}
    \centering
    \includegraphics[width=0.98\columnwidth]{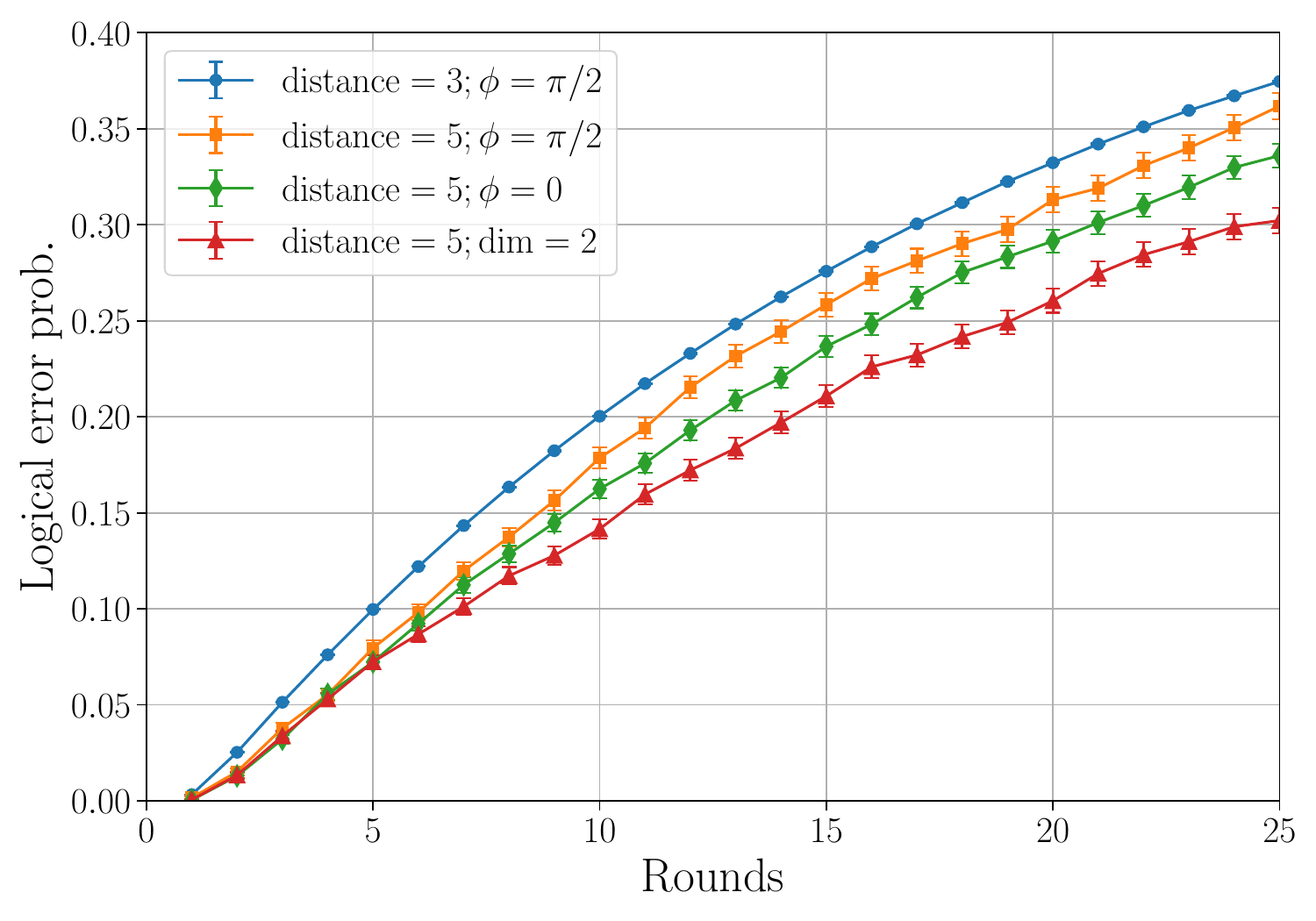}
    \caption{\textbf{Distance 5 surface code simulations.} Here we compare various surface code simulations using different parameters, plotting the logical error probability in a memory experiment. The three data sets with $\phi$ in the legend are qutrit simulations utilizing the STA, where $\phi$ is the $|21\rangle$ phase angle of the CZ gate (see main text). The simulations otherwise were run with the same physical parameters as in App.~\ref{sec:STA-physical}. See App.~\ref{sec:simulation} for additional implementation details. The red triangles correspond to a purely qubit simulation (no leakage). Distance 5 data from 5000 samples.}
    \label{fig:dist_5_comparison}
\end{figure}

\section{Discussion}

In this work we have introduced a Subspace Twirling Approximation (STA): a way to approximate quantum channels as incoherent with respect to a given collection of subspaces. We have detailed how this approximation can be used to model leakage while retaining only a state-vector dimension scaling as $O(2^n)$. This enables a fully quantum simulation of a distance 5 surface code with leakage, which nominally requires the equivalent of $41\sim \log_2 3^{26}$ qubits in memory. The STA reduces this to just 26 qubits which can be handled with relative ease; the distance 5 simulation with leakage in Fig.~\ref{fig:dist_5_comparison} required $\sim 2.5$ minutes of simulation time per sample per round, on a single Intel Xeon Cascade Lake 2.7GHz core. 

Since the STA still requires exponential resources, it is a weaker approximation than the Generalized Pauli Twirling Approximation (GPTA). The latter was used to efficiently model the effect of leakage in Clifford simulations~\cite{scaling_SC_Google_2022}. We note that the GPTA already treats leakage as an incoherent process. Thus, if we are to accurately simulate error correction at the scale where only Clifford-type simulations are feasible, we must first justify the STA for leakage; if the STA is not accurate, certainly the GPTA will not be, as it is a strictly stronger approximation. In other words, agreement between the GPTA with the STA is a necessary (but not sufficient) condition for accurate GPTA simulations. 

Our simulations suggest that, within the context of common error correction experiments, the STA is an accurate approximation for the common leakage mechanisms present in current transmon qubit-based architectures. In the atypical case where leakage is a purely coherent/unitary process, we have shown that an effective thermal model can still be used to accurately predict error metrics such as detection event fractions, leakage population, and logical error. Additionally, in App.~\ref{sec:STA-physical} we show that the STA is accurate for a more realistic combination of unitary and thermal leakage models, without the need for an effective model. Furthermore, we expect that the additional of Leakage Reduction Units (LRUs)~\cite{google-removing-leakage-2021,google-leakage-2022, Terhal-2021-LRU} after every error correction round may further improve the accuracy of the STA (while simultaneously mitigating the impact of leakage). These operations are engineered to cause transitions from the leakage subspace into the computational subspace of a given qubit, while leaving the computational subspace unchanged. Since this directly produces an incoherent state, any coherent effects impacting leakage accumulation are truncated to just one round of error correction.

As an alternative to the effective thermal model, one could also use the STA directly on the coherent leakage error channel to produce an effective incoherent model. While perhaps more cumbersome, this could likewise be fit to observed leakage populations over time. We caution, though, that fitting such a model to direct experimental characterization of the CZ gate may \emph{not} produce the most accurate effective model. This is because it would ignore the coherent growth of leakage in the context of multiple rounds of CZ gates. As an illustration of this, in App.~\ref{sec:STA-additional}, Figure~\ref{fig:leakage-eta}, we consider the effect of phase accumulation on state $\ket{2}$ between CZ gates. This phase has a significant impact on the growth of leakage populations over time, but because it is dependent on the gates after the CZ, it cannot extracted from direct characterization such as process tomography~\cite{bialczak2010quantum, mohseni2008quantum}. As such inferring a accurate effective incoherent model requires context-aware characterization of CZ leakage~\cite{CAFE}.

In addition to justifying the STA, our simulations  provide several physical insights regarding leakage in the surface code. First, as noted previously~\cite{leakage-detection}, the phase of the coherent leakage transitions $\ket{11}\leftrightarrow\ket{02}$ does not observably affect the logical error probability or leakage accumulation (see App.~\ref{sec:STA-additional}). On the other hand, the \emph{controlled phase} in the CZ gate when the data qubit is leaked, $\sim \exp(i\phi Z/2) \otimes \ketbrad{2}$, does have a significant impact on the observed logical error. We can see this clearly in Fig.~\ref{fig:dist_5_comparison} which uses experimentally realistic parameters in the distance 5 surface code. We verified that in this case the leakage induced logical error is minimized at $\phi= 0, \pi$, and maximized at $\phi = \pi/2$ (the logical error rates for the curves in Fig.~\ref{fig:dist_5_comparison} are 2.81\%, 2.61\%, 2.32\%, and 1.94\%, from top to bottom respectively). As discussed in Sec.~\ref{sec:coherence-decay}, this phase is directly related to the frequency of leakage-induced detection events, and it determines the rate at which the stabilizer measurements cause the coherence of leakage to decay. In practice, the phase $\phi$ is sensitive to the details of the gate implementation (e.g., the frequency trajectory of each qubit, coupling strengths, and the gate timings). Modifying the CZ gate to also fix this phase could therefore be used to partially mitigate the effect of leakage. Another mitigation strategy is motivated by the simulations seen in appendix Fig.~\ref{fig:leakage-eta}, which show that coherent leakage accumulation depends strongly on the the transmon nonlinearity (and therefore the phase accumulated by state $\ket{2}$ between identical CZ gates). This suggests that coherent leakage mechanisms between rounds of error correction could be made to destructively interfere by applying an appropriate phase between the leakage and computational subspace.

We conclude by mentioning some possible applications of the STA beyond the study of leakage in error correction. Overall, the assumption of incoherence gives a prescription for generating effective classical models that still incorporate long-time correlations. These models are described using classical registers requiring a simulation cost only linear in the number of registers. For example, they can be used to efficiently treat environmental degrees of freedom that strongly couple to the qubits, such as strongly coupled two-level systems or readout resonators. Importantly, the STA can be used to map a coherent interaction with these environmental modes into one that is incoherent with respect to the environmental states. (The validity of the approximation will depend on the system of interest.) The tools we have introduced provide another paradigm in which to perform simulations of leakage, such as approximations based on the stabilizer formalism \cite{scaling_SC_Google_2022}, or tensor network methods in thin surface codes \cite{manabe_efficient_2023}. Going forward we aim to leverage this technique to understand the impact of other error mechanisms on error correction fidelity.

\section{Data Availability Statement}
Data, additional details, etc., available upon reasonable request.

\section{Acknowledgements}
We thank Namit Anand, Michael Newman, and Agustin Di Paolo for reading the manuscript and providing helpful comments.

J.M. acknowledges the support from NASA Ames Research Center, funding under NASA-Google SAA2 403512 and 403706, and support from the NASA Advanced Supercomputing Division for providing access to the NASA HPC systems.
J.M. is additionally grateful for support from NASA Academic Mission Services (NAMS), Contract No.~NNA16BD14C.

\bibliography{refs}

\begin{thebibliography}{10}
\bibitem{fn-dq-leak}
{We focus on data qubits because the measure qubits used to effect the
  stabilizer measurement are typically measured at every round of error
  correction. Even if the measure qubits are not reset to the zero state each
  round, this strong projective measurement prevents the leakage amplitude from
  accumulating coherently}.
\end{thebibliography}

\appendix

\section{Numerical Implementation of Subspace Twirling Approximation}
\label{sec:STA-numerics}
Here we describe how the STA is numerically implemented and integrated into the quantum trajectories simulations. We consider the typical case of a quantum operation acting on a well-defined number of qubit degrees of freedom. At the end of the section we comment on generalizations to operations that create or destroy qubits, as well as operations whose sampled Kraus operator (prior to applying STA) is stored as a classical register.

The STA is implemented using the quantum and classical operations discussed in Section~\ref{sec:channel-rep}. We consider a quantum channel $\mathcal E$ acting on a subset $Q$ of the qubit degrees of freedom. We can describe the local Hilbert space of these systems as a tensor product
\begin{equation}
    \mathcal H = \bigotimes_{q\in Q} \mathcal C_q \oplus \mathcal L_q
\end{equation}
where $\mathcal C_q$ denotes the local computational subspace of each qubit $q$ and $\mathcal L_q$ the leakage subspace. (For simplicity we assume this space is one dimensional and spanned by just the leakage state $\ket{2}$, though generalizing to more states is straightforward.) The local Hilbert space $\mathcal H$ can then be rewritten as a direct sum
\begin{equation}
    \label{eq:HS-direct-sum}
    \mathcal H = \bigoplus_{\bar r} \mathcal P_{\bar r}
\end{equation}
\begin{equation}
    \label{eq:local-subspaces}
    \mathcal P_{\bar r} =\bigotimes_{q\in Q} \mathcal P_{r_q} 
\end{equation}
where $\bar r$ is a vector of length $|Q|$ with symbolic values $r_{q}\in \{c, 2\}$, and $\mathcal P_{r_q} \in \{\mathcal C, \mathcal L \}$.
We numerically represent the basis for each $\mathcal P_{\bar r}$ as
$$X_{\bar r} = X_{r_1} \otimes X_{r_2} \otimes \,... X_{r_{|Q|}}.$$
Written as a matrix, the operator $X_{r_q}$ has shape $3\times 2$ (for $r_q=c$) or $3\times 1$ (for $r_q=2$).
Under the STA, the Kraus operators $K_j$ of channel $\mathcal E$ are transformed to new Kraus operators as in Eqs.~\eqref{eq:KrausCrossSTA} and~\eqref{eq:KrausDiagSTA}. The new operators are either block diagonal with respect to $\mathcal P_{\bar r}$ or are off-diagonal with exactly one non-zero block. Each Kraus operator $K_j$ under the transformation can therefore be mapped to its blocks:
\begin{equation}
\label{eq:kraus-blocks}
    K_{j,{\bar s,\bar t}} = X_{\bar s}^\dagger K_j X_{\bar t}.
\end{equation}
We use these blocks to express $\mathcal E_{STA}$ as a combined quantum and classical channel. This has the following Kraus operators:
\begin{equation}
    \l\{K_{j,{\bar s,\bar t}}  \otimes \ketbradt{\bar s}{\bar t}\r\}_{j, \bar s \neq \bar t} 
        \bigcup \l\{\sum_{\bar t} K_{j,{\bar t,\bar t}}  \otimes  \ketbrad{\bar t}\r\}_{j} .
\end{equation}
The first type of Kraus operators denote a transition from $\mathcal P_{\bar t}$ to $\mathcal P_{\bar s}$ (operators of type II in Section~\ref{sec:STA}), while the second (type I) represent applying Kraus operator with index $j$ and remaining in a given subspace. For an initial state that is \emph{incoherent} with respect to the subspaces $\mathcal P_{\bar s}$, the action of $\mathcal E_{STA}$ is unchanged if we replace each of the ``block-diagonal'' Kraus operators with multiple ``single subspace'' operators $K_{j, \bar t, \bar t}\otimes \ketbrad{\bar t}$. This results in a simpler set of Kraus operators,
\begin{equation}
   \label{eq:krausops-STA-effective}
   \l\{K_{j,{\bar s,\bar t}}  \otimes \ketbradt{\bar s}{\bar t}\r\}_{j, \bar s , \bar t} .
\end{equation}
This representation is equivalent for incoherent simulations but more convenient for book-keeping purposes. The structure denoted by equation~\eqref{eq:krausops-STA-effective} is stored as a hashmap from pairs $(\bar s, \bar t)$ to the corresponding blocks $[K_{j_1, \bar s,\bar t}, K_{j_2,\bar s,\bar t}, ...]$. For each ``transition'' $\bar t \rightarrow \bar s$ the sequence runs through only a subspace of all $j$, since we only keep the blocks which are non-zero (up to some truncation tolerance). Accordingly, we also maintain the sequence of the indices $[j_1, j_2, ...]$.

The Kraus operator representation in Eq.~\eqref{eq:krausops-STA-effective} is not in one of the standard forms described in Sect.~\ref{sec:channel-rep}. (Specifically, the standard form does not allow a Kraus operator to act non-trivially on both existing quantum and classical degrees of freedom.) To map it to the desired form, we break it up into two separate channels.
The first channel is written as
$$ \{ K_{j,{\bar s,\bar t}}  \otimes \ketbrad{\bar t}\otimes\ket{\bar s} \}_{j, \bar s, \bar t}.$$
Writing the (quantum and classical) state \emph{prior} to application of the channel as
$$\ket{\psi}\otimes \ket{\bar t}\,,$$
the first channel represents the application of the quantum Kraus operators $K_{j,\bar s, \bar t}$, conditioned on the current leakage subspace being $\mathcal P_{\bar t}$. Additionally, it creates a set of classical registers to denote which subspace $\mathcal P_{\bar s}$ to which the qubits transitioned. 
In a trajectories simulation, this corresponds to a lookup of which subspace $\mathcal P_{\bar t}$ the system is currently in, followed by a sampling of the Kraus operators $K_{j,{\bar s,\bar t}}$ (for fixed $\bar t$) on the quantum state vector.
After this channel is applied, the ``recorded'' classical registers (with values $\bar s $) are consumed to map the original classical registers to their new values. This is denoted by the second channel (acting only on the classical registers),
$$\{ \ketbradt{\bar s}{\bar t}\otimes \bra{\bar s}\}_{\bar s, \bar t}.$$
(This ``classical'' channel acts trivially on the quantum degrees of freedom.)

The prescription we give above can be directly generalized in several ways.
First, the sampled Kraus operator index $j$ (prior to applying the STA) can also be recorded in the above decomposition, through the modification $\ket{\bar s} \rightarrow \ket{j}\otimes \ket{ \bar s}$.  This can be used to represent a measurement outcome or capture some other physical process that models an interaction between quantum and classical degrees of freedom.
Second, the original Kraus operators $K_j$ need not be square matrices. For example, consider the destructive measurement of a qubit and subsequent recording of the outcome to a classical register, which has Kraus operators $\{\bra{j} \otimes \ket{j}\}_j $. Under our STA representation this would correspond to having a single initial subspace register $t$ and zero output subspace registers.
Finally, the STA is not limited to the case of computational and leakage subspaces. It should apply in any situation where we can decompose the full system Hilbert space as in Eqs.~\eqref{eq:HS-direct-sum} and~\eqref{eq:local-subspaces}. The computational usefulness of this approximation will depend on whether the noise channels $\mathcal{E}$ have bounded locality with respect to the tensor product decomposition of $\mathcal H$, as well as whether it is accurate in the given physical context.

\section{\label{sec:reordering-trick}Reordering trick}

Here we provide a deterministic algorithm for implementing the ``reordering trick'' first introduced in Ref.~\cite{obrien-sc-dm-sim}. This technique allows for a reordering of the circuit operations (through commutation of channels acting on different qubits) to minimize the number of measure qubits required in memory during a quantum trajectory. For repetition and surface-code circuits, we find that \emph{at most one} measure qubit is ever required in memory. Below is a short summary of how this is carried out.

The reordering trick first requires us to transform the measurement and reset operations. Measurements are converted into \emph{destructive measurements} (see Sect.~\ref{sec:channel-rep}). Each measurement causes the number of degrees of freedom to decrease by one, and correspondingly lowers the size of the quantum state vector.
Measurements on a given measure qubit are then followed by a \emph{creative reset}. Each such operation increases the number of degrees of freedom by one. If physical reset operations are used in the circuit, the creative resets deterministically prepares the measure qubit in the target state $\ket{0}$. Otherwise, the measure qubit is prepared in the standard basis state in which it was last measured.

\begin{figure}
    \centering
    \includegraphics[width=0.95\columnwidth]{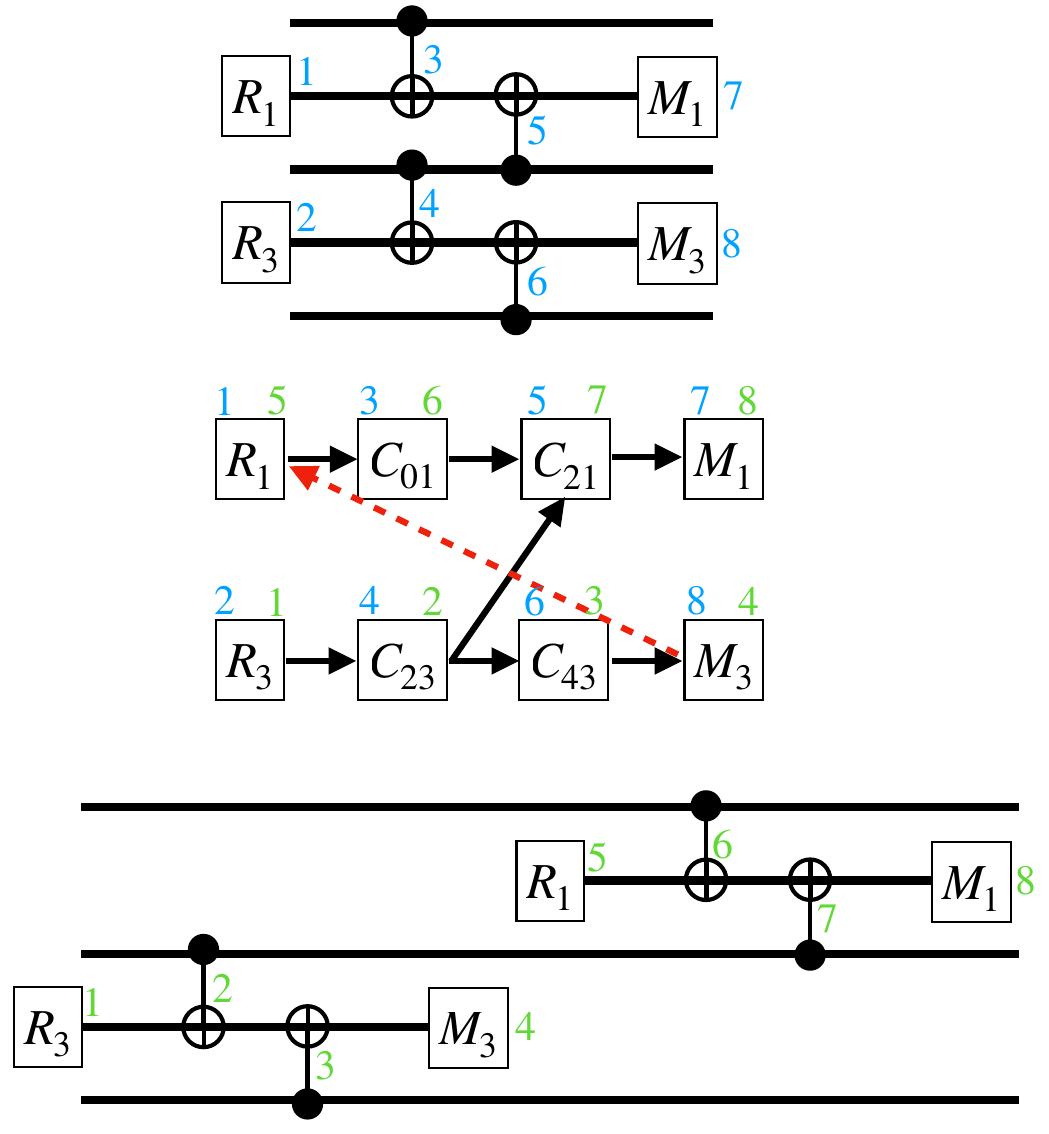}
    \caption{\textbf{Example of topological reordering.} (Top) Single round of a repetition code (distance 3), which nominally retains 5 qubits in memory.
    The measure qubits (1,3) begin with a creative reset and end with a destructive measurement. The data qubits are on the other three lines. 
    The index (blue) to the right of each operation indicates the canonical execution order of the operations, where in a given time step (vertical slice), the operation that acts on the lowest qubit index is applied first (though, as operations in a time step commute, this is actually arbitrary).
    (Middle) The DAG, constructed as in the main text, where the dashed (red) arrow is a new edge that gets added, based on the fact the $C_{21}$ must occur after $C_{23}$, where $C_{ij}$ means a $CX$ gate between qubits $i,j$.  Performing a topological reordering of this DAG results in a circuit equivalent to the bottom one. The top left (blue) indices are the order the operation appears before reordering, and the top right (green) after reordering. (Bottom) The reordered circuit. Notice that in this circuit at most 4 qubits are in memory at any instance of time. In particular, adding the dashed edge enforces that $M_3$ occurs before $R_1$, even though that is not a requirement in the initial circuit. In this case, the topological ordering is unique: $[R_3, C_{23}, C_{43}, M_3, R_1, C_{01}, C_{21}, M_1]$. This order is shown by the index (green) to the right of each operation.}
    \label{fig:reorder-example}
\end{figure}

Next, the (noisy) error correction circuit, considered as a sequence of quantum channels $\mathcal E_1, \mathcal E_2, ...$, is decomposed into a graph. Each node $j$ of the graph is associated with a single quantum channel $\mathcal E_j$. A directed edge is assigned from node $j$ to node $k$ whenever 
    \begin{enumerate}
        \item $j < k$ 
        \item There exists a degree of freedom $q$ acted on nontrivially by both $\mathcal E_j$ and $\mathcal E_k$.
        \item There exists no $l$ such that $j < l < k$ and $\mathcal E_l$ acts nontrivially on $q$ (i.e., from the perspective of $q$, channel $\mathcal E_k$ is the next channel applied to it after $\mathcal E_j$).
    \end{enumerate}
Observe that any ordering of the nodes (channels) $\mathcal E_{s_1}, \mathcal E_{s_2}, ...$ is physically equivalent to the original circuit as long as the order implied by the directed edges is conserved. This means that if $\mathcal E_{s_a}$ and $\mathcal E_{s_b}$ both act nontrivially on $q$ and $a<b$, then a directed path exists from $s_a$ to $s_b$ (in this specific case, this implies that $s_a < s_b$). Such an ordering is called a \emph{topological ordering} of the directed graph.
An example of the directed acyclic graph (DAG) after this step is shown in Fig.~\ref{fig:reorder-example} (middle), excluding the dashed (red) edge, which gets added below.

To enable the desired memory-efficient ordering, we add additional directed edges to the graph. (Since edges are only added and not removed, an ordering that is consistent with the new graph will also be consistent with the original.) We use the following rule: In a single error correction round, if there exists $j<k$ such that 
    \begin{enumerate}
        \item Channel $\mathcal E_{j}$ acts on measure qubit $q_m$ and data qubit $q_d$, and
        \item Channel $\mathcal E_{k}$ acts on measure qubit $q_m'\neq q_m$ and the same data qubit $q_d$
    \end{enumerate} 
then we know that information about $q_m$ can be transmitted to $q_m'$ through $q_d$. In this case, we add an edge between destructive measurement of $q_m$ and the creative reset (following measurement in the same round) of $q_m'$. A topological ordering of this graph, while still physically equivalent to the original circuit ordering, ensures that qubit $q_m$ is destroyed before qubit $q_m'$ is created. We note that for the circuits we considered throughout this paper, the additional edges never cause a cycle in the graph. The algorithm is straightforward to generalize to this case: identify the cycle-forming measure qubits as the same qubit (though only for the purpose of reordering the operations). The measure qubits associated with a given cycle will need to be simultaneously included in memory during simulation. 
In Fig.~\ref{fig:reorder-example}, the dashed (red) edge in the DAG is the result of this final step. In particular, the data qubit 2 is acted on by $C_{23}$ then $C_{21}$, so we add an edge between the measurement of 3 and reset of 1. Performing the topological reordering then results in the more memory efficient circuit shown, as it guarantees qubit 3 is removed from memory before qubit 1 is added.

\section{Simulation details \label{sec:simulation}}
\begin{figure}
    \centering
    \includegraphics[width=0.98\columnwidth]{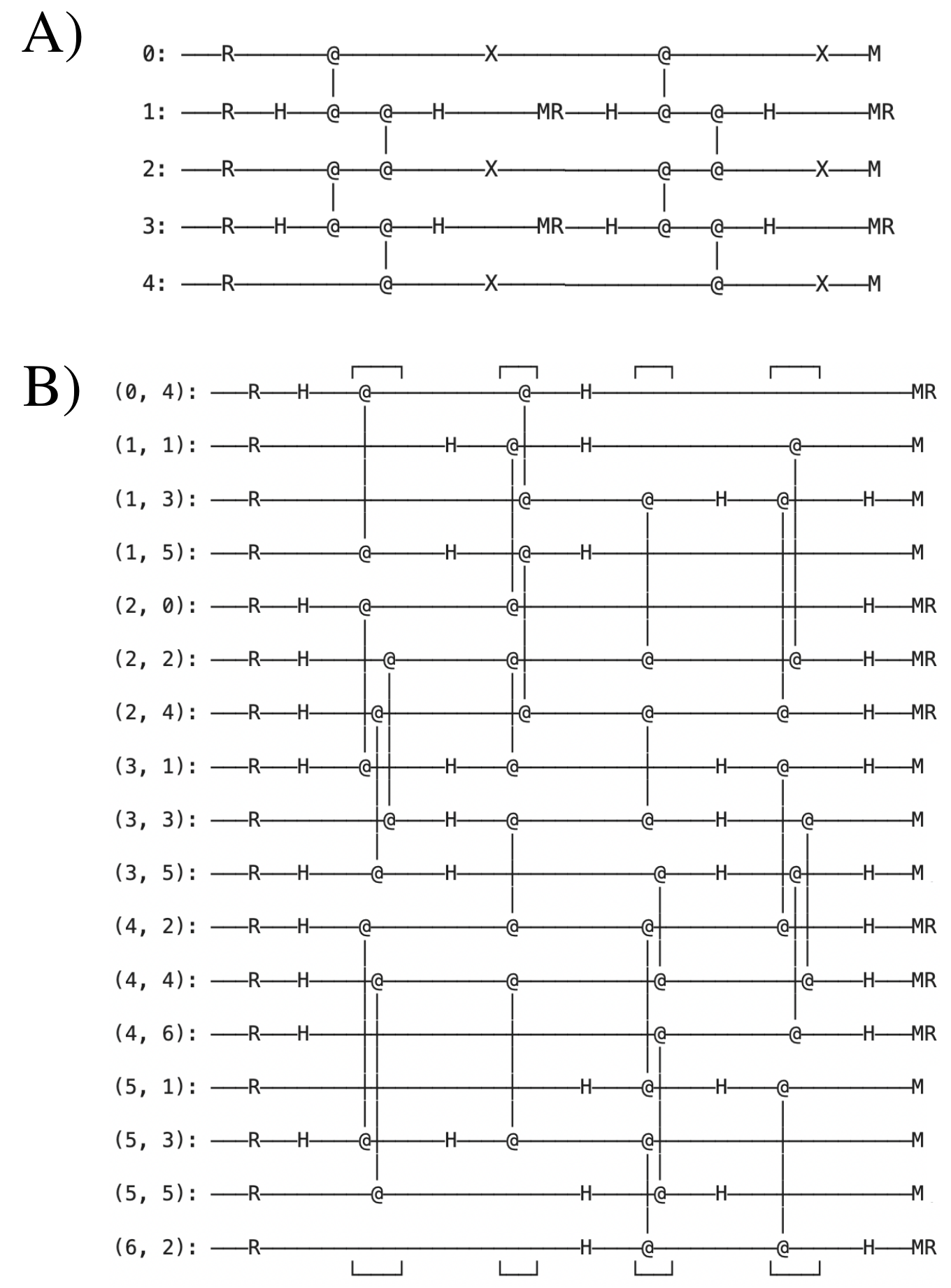}
    \caption{\textbf{A)} Two rounds of distance 3 Z basis repetition code (even integers label data qubits).
    We introduce a logical flip ($X$ gates) at the end of each round to symmetrize over the dominant $T_1$ errors (to avoid the system decaying to the all $|0\rangle$ state). We typically begin in a random bit-string configuration (by applying $X$ gates on appropriate qubits after the initial reset `R').
    \textbf{B)} Single round of distance 3 rotated surface code (odd coordinates label data qubits). Even coordinates that sum to $0\mod 4$ ($2\mod 4)$ are $Z$ ($X$) stabilizers.
    In the SC we typically begin in the all $|0\rangle$ state.
    `H' indicates a Hadamard gate, and `$@-@$' is a CZ gate.
    Measure qubits are reset to $|0\rangle$ at the beginning of each round (`R'), and measured at the end of each round (`M'). At the end of the simulation, the data qubits are also measured. Hadamard, $X$ and CZ gates are set in simulation to $25$ns, reset $600$ns, and measurement $300$ns. Circuits displayed diagrammatically using \texttt{Cirq} \cite{cirq_developers_2022_6599601}.}
    \label{fig:rc_sc_schematic}
\end{figure}

\begin{figure}
    \centering
    \includegraphics[width=0.98\columnwidth]{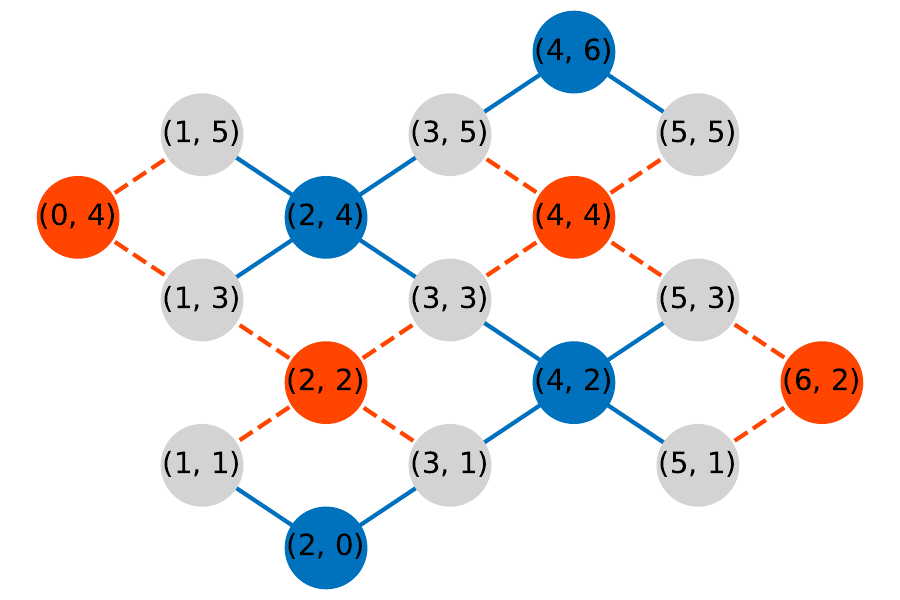}
    \caption{\textbf{Schematic of distance 3 rotated surface code used in this work}. Data qubits have odd coordinates. Even qubits with dash lines (red) measure $Z$ stabilizers, and solid lines (blue) are $X$ stabilizers.}
    \label{fig:SC_schematic}
\end{figure}

Here we outline details of the simulations used in the main text. In Fig.~\ref{fig:rc_sc_schematic} we show distance 3 versions of the repetition and rotated surface code used in our simulations. We generate the circuits using \texttt{Stim} \cite{gidney2021stim}. 
We provide a schematic layout of the distance 3 SC in Fig.~\ref{fig:SC_schematic}. 
Using the reordering trick (App.~\ref{sec:reordering-trick}) means a distance $d$ surface code (repetition code) requires at most $d^2+1$ ($d+1$) qubits to be stored in memory, instead of $2d^2-1$ ($2d-1$).

Our simulations include two classes of noise. We give a high level description here, and provide more details below. I) thermal models (including leakage transitions) and dephasing that are applied after each gate (or idle) as a Lindblad equation evolved over the appropriate duration. II) coherent leakage via the CZ gate with transitions $|11\rangle \leftrightarrow |02\rangle$.
In the case of leakage, any other unitary (such as Hadamard or Pauli unitaries) acts trivially on the additional levels.

Here we discuss the thermal and dephasing models used in our simulations. Dephasing, heating and cooling on single qubits is achieved by solving a (time-independent) Lindblad master equation, which takes as input the rates for various transitions. In general we consider a Lindblad master equation with three Lindblad operators of the form (in a qutrit system):
\begin{equation}
\begin{split}
    & L_{heat} = \left(\begin{array}{ccc}
       0  & 0 & 0  \\
       \sqrt{\gamma_{10}}  & 0 & 0  \\
       0  & \sqrt{\gamma_{21}} & 0 
    \end{array}\right);\;     
    L_{cool} = \left(\begin{array}{ccc}
       0  & \sqrt{\gamma_{01}} & 0  \\
       0  & 0 & \sqrt{\gamma_{12}}  \\
       0  & 0 & 0 
    \end{array}\right); \\     
    & L_{dephase} = \left(\begin{array}{ccc}
       0  & 0 & 0  \\
       0  & \sqrt{\gamma_{11}} & 0  \\
       0  & 0 & \sqrt{\gamma_{22}} 
    \end{array}\right).
\end{split}
\label{eq:lindblad_ops}
\end{equation}
For $i \neq j$, $\gamma_{ji}$  defines the transition rate (per unit time) from $\ket{i}\rightarrow\ket{j}$. The dephasing rates are typically controlled by a single time-scale, $T_\phi$, i.e., $L_{dephase} = \sqrt{2/T_\phi}n$ where $n=a^\dag a$ is the number operator. In the present work, the heating is implemented as $L_{heat} = \sqrt{{1}/{T_h}}a^\dag$, for a characteristic time-scale $T_h$. In this case, the $|2\rangle \langle 1|$ matrix element is $\sqrt{2/T_h}$, resulting in $1\rightarrow 2$ transition time of $\gamma_{21}^{-1}=T_h/2$.
The $T_1$ time-scale defines $\gamma_{01} = 1/T_1$. The mean leakage lifetime $T_L$ is controlled by $\gamma_{12}=1/T_L$. 
The dynamical equation is then of the form $\dot{\rho} = \mathcal{L}\rho = ( \mathcal{L}_{heat} + \mathcal{L}_{cool} + \mathcal{L}_{dephase}) \rho$, where
\begin{equation}
    \mathcal{L}_{heat}X = L_{heat}XL_{heat}^\dag - \{L_{heat}^\dag L_{heat}, X\}
\end{equation}
etc.

The simulation is performed by extracting Kraus operators from the solution of the dynamics, $\exp(t \mathcal{L})$ (e.g., see \cite{wood_tensor_2015}). These Kraus operators are sampled as discussed in the main text (\ref{sec:traj-sims}). The time $t$ is determined by gate/idle times, and the model is applied on each qubit, after each gate or idle (where other qubits are active). The unitary gate times have 25ns duration, with 600ns reset and 300ns measurement (in line with experimental capabilities). As a result, each round of the surface and repetition code has approximately 1$\mu$s duration. 

We now describe the qutrit ($9\times 9$) CZ gate, which is composed of two parts; a diagonal matrix of phases, and a transition (sub)matrix. We allow a non-trivial phase to accumulate on the $\ket{12}, \ket{02}$ states by introducing phase factors  $\phi_{12}, \phi_{02}$ in the diagonal unitary. As we discuss in Sec.~\ref{sec:coherence-decay} only the difference $\phi=\phi_{02}-\phi_{12}$ is observable, and this can have a dramatic influence on the decoding of leakage errors.  Unless otherwise stated we set $\phi=\pi/2$. All other phases are trivial (0), apart from $|11\rangle$ which has a $\pi$ phase.
The qubit nonlinearity $\eta$ is also encoded, where appropriate (leaked) states acquire an additional phase $\eta t$, where $t$ is the duration of the gate.
The CZ gates we model can further introduce coherent transitions $|11\rangle \leftrightarrow |02\rangle$, where the qubit at the higher frequency transitions to $\ket{2}$ (in our simulations this is always a data qubit). We model this in simulations with a unitary submatrix that multiplies the diagonal `phase' matrix (in the $|11\rangle, |02\rangle$ subspace):
\begin{equation}
    \left(\begin{array}{cc}
       \sqrt{1-p}  & -e^{i\phi_{11,02}}\sqrt{p} \\
       e^{-i\phi_{11,02}}\sqrt{p}  & \sqrt{1-p}
    \end{array}\right).
\label{eq:leak_matrix}
\end{equation}
We call $p$ the `transition probability'. The `transition phase' $\phi_{11,02}$ is a free parameter for simulations, which we find has little effect on our results (e.g. Fig.~\ref{fig:leakage-transition-phase}), and thus we typically set this to 0.

Simulations are conducted using our in-house Kraus operator sampler, \texttt{kraus-sim}, which updates the state vector according to the prescritpion outlined in Sect.~\ref{sec:traj-sims} (further documentation available in Ref.~\cite{scaling_SC_Google_2022}). The backend of the simulator uses either \texttt{NumPy} or \texttt{qsim} \cite{qsim_2020_4023103}.
In order to decode our error correction memory experiments we use \texttt{Stim} \cite{gidney2021stim} in conjunction with \texttt{PyMatching} \cite{higgott2021pymatching}, which runs a minimum-weight perfect matching algorithm. For the detector error model, we use a depolarizing model after each gate, with depolarizing probability 0.001 (though we find the decoding success is fairly robust to changes in this value). The extraction of logical error rates is described below in App.~\ref{sec:error-rate}.
In order to perform the decoding with leakage present, any measurement result $2$ is mapped randomly to 0 or 1 (with equal probability).

\section{Logical error rate \label{sec:error-rate}}
\begin{figure}
    \centering
    \includegraphics[width=0.98\columnwidth]{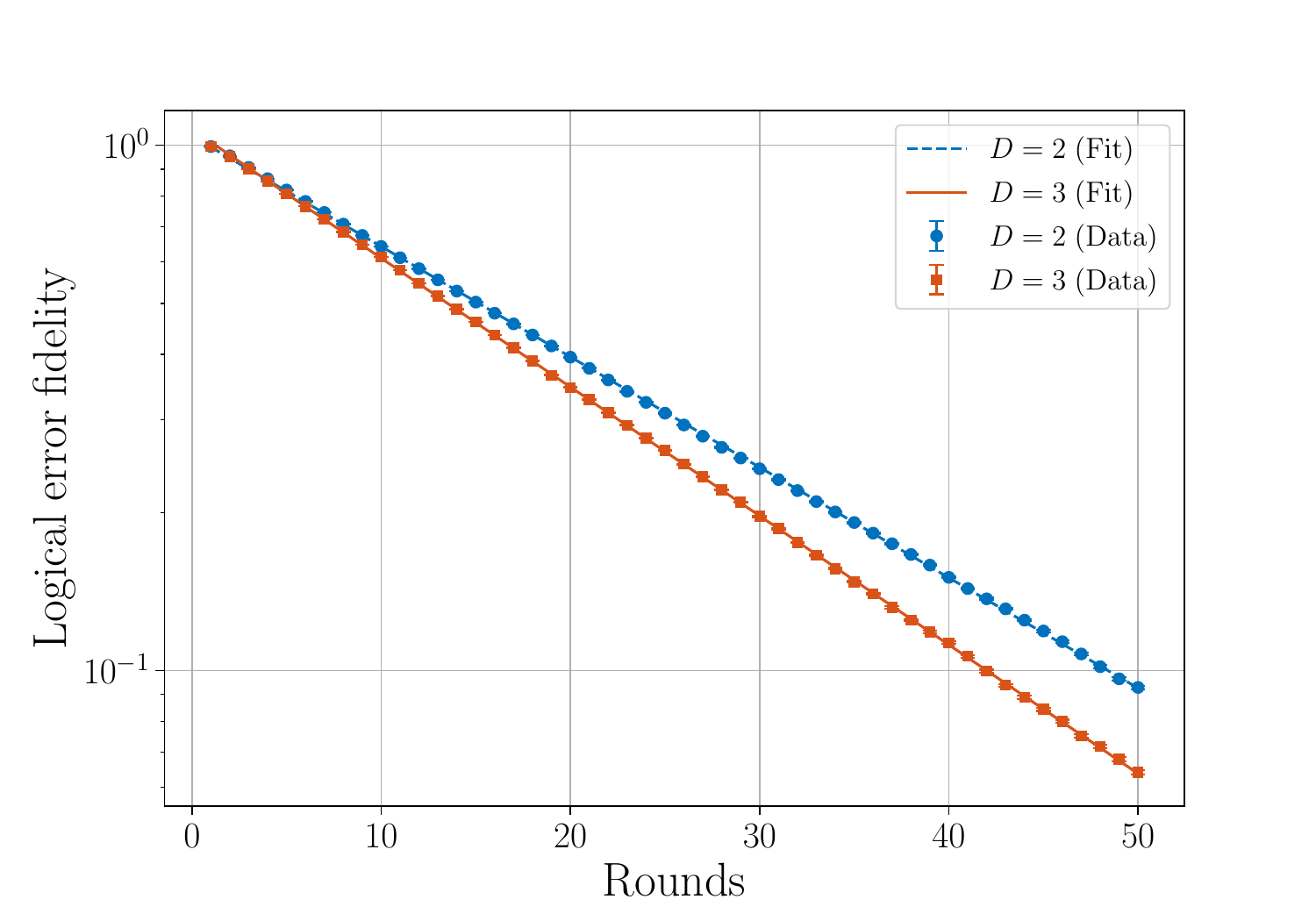}
    \caption{Logical error fidelity fit for the main text data from Fig.~\ref{fig:cz-logical-err}. $D=2$ (qubit) corresponds the the simulation with no leakage, and $D=3$ (qutrit) has leakage from the CZ gate. The $D=2$ case has fitting parameters [Eq.~\eqref{eq:LEF}] of $A=1.04, \epsilon_L=0.0236$, and $D=3$ has $A=1.07, \epsilon_L=0.0275$. Data points are the circles/squares, and lines the least squares fit.}
    \label{fig:LEF_fit}
\end{figure}

Here we discuss the extraction of the logical error rate (LER) from simulation/experimental data, using the model of Ref.~\cite{obrien-sc-dm-sim}. In particular, noting that only an odd number of physical errors contribute to a logical error, one can show the error probability is governed by 
\begin{equation}
    P_L(k) = \frac{1}{2}(1 - A(1-2\epsilon_L)^{k}),
\end{equation}
where $k$ is the number of rounds, and $\epsilon_L$ the logical error rate. Here $A\approx 1$ is introduced ad hoc as an additional fitting parameter (typically $A\in [1, 1.08]$ in our simulations). In practice we perform a straight line fit to the logarithm of the logical error fidelity
\begin{equation}
    F_L(k) = 1 - 2P_L(k) = A (1 - 2\epsilon_L)^{k}.
    \label{eq:LEF}
\end{equation}
In Fig.~\ref{fig:LEF_fit} we show an example of such a fit using data from the main text. Notice that the case with leakage has a larger value of the fitting parameter $A$. We find this is typically true in such simulations with leakage, likely the result of the equation \eqref{eq:LEF} not taking into account the dynamics of leakage (only after a sufficient number of rounds does leakage saturate, e.g.~see Fig.~\ref{fig:leakage-eta}). 
Nevertheless, we can see, at least at longer times, the equation provides a good approximation to the decay of fidelity.

\section{Additional STA simulation data \label{sec:STA-additional}}

\begin{figure*}
    \centering
    \includegraphics[width=1.25\columnwidth]{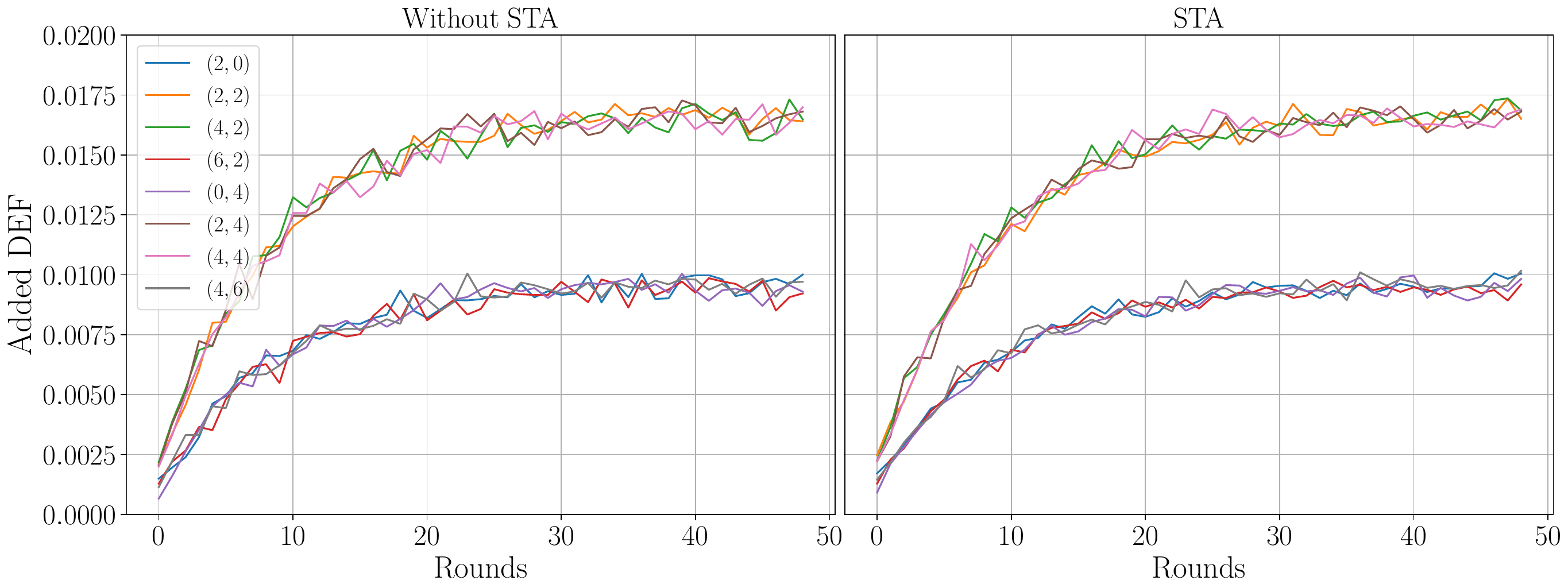}
    \caption{\textbf{Detection event fraction (DEF) under STA for thermal model.} Distance 3 surface code where all leakage comes from `thermal heating' as discussed in the main text, the same data as in Fig.~\ref{fig:thermal-logical-err}.}
    \label{fig:def-thermal}
\end{figure*}

Here we provide supplemental plots and data pertaining to the validation of the Subspace Twirling Approximation,  Sect.~\ref{sec:STA_validation}. 
First, we show that the detection event statistics relating to the main text `thermal heating model', Fig.~\ref{fig:thermal-logical-err}, are consistent in the model with and without the STA. A detection event is defined as a particular stabilizer value flipping from one round to the next (due to a physical error) \cite{chen_exponential_2021}. We compute the detection event fraction (DEF), which is the probability a given measure qubit (stabilizer) registers a detection event, in a given round. It is more instructive to plot the \emph{added} DEF due to leakage, obtained by considering the difference in DEFs between simulations with and without leakage. In Fig.~\ref{fig:def-thermal}, we see the full qutrit simulation (left panel) statistics match fairly well to the STA (two-level) simulation (right panel), with fluctuations likely caused by finite sampling.

Next we consider the second leakage model of interest, with strong CZ leakage, and no thermal heating, as in Sect.~\ref{sec:STA-validation-coherent}.
As a reminder, here all leakage comes from the diabatic CZ gate, via transitions $\ket{11}\rightarrow \sqrt{1-p}\ket{11} + e^{i\phi_{11,02}}\sqrt{p}\ket{02}$, where the first qubit in the pair (the one that leaks) is always a data qubit. We set the leakage probability to a relatively large value, $p=2.4\times 10^{-3}$.

\begin{figure*}
    \centering
    \includegraphics[width=1.75\columnwidth]{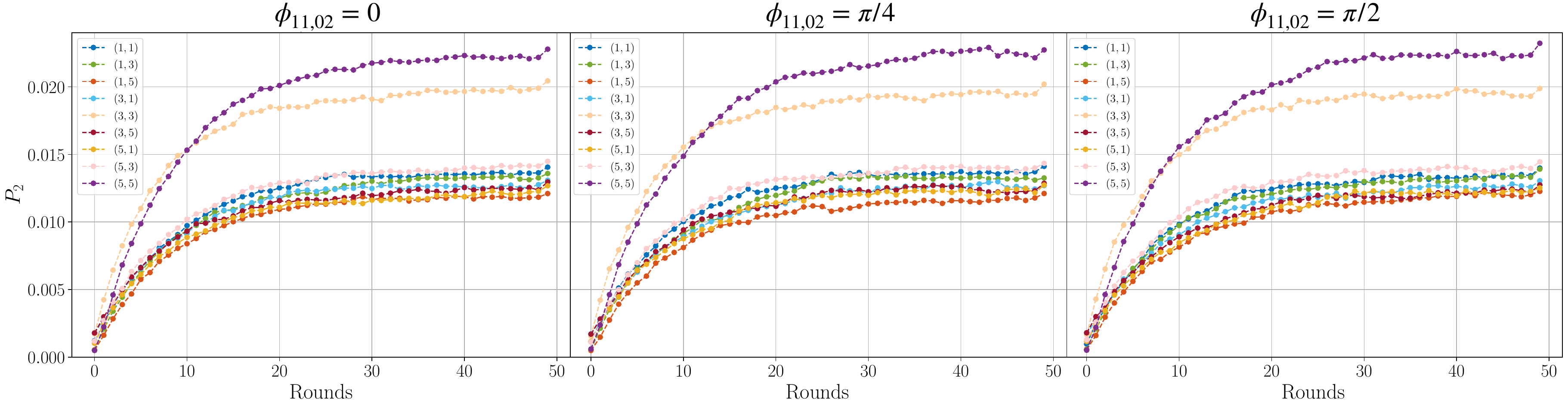}
    \caption{\textbf{Leakage populations with varying CZ transition phase.} For the distance 3 surface code, with strong CZ leakage (cf.~Fig.~\ref{fig:cz-logical-err}), we plot the two state population for all data qubits, where from left to right the CZ transition phase is $\phi_{11,02} = 0, \pi/4, \pi/2$. The CZ transition phase between $\ket{11}, \ket{02}$ is applied during $\ket{11}\rightarrow \sqrt{1-p}\ket{11} + e^{i\phi_{11,02}}\sqrt{p}\ket{02}$. Note, these simulations were preformed without invoking the STA, taking at least 250k trajectories. The non-linearity is set to $\eta=0.2$GHz.}
    \label{fig:leakage-transition-phase}
\end{figure*}

\begin{figure}
    \centering
    \includegraphics[width=0.9\columnwidth]{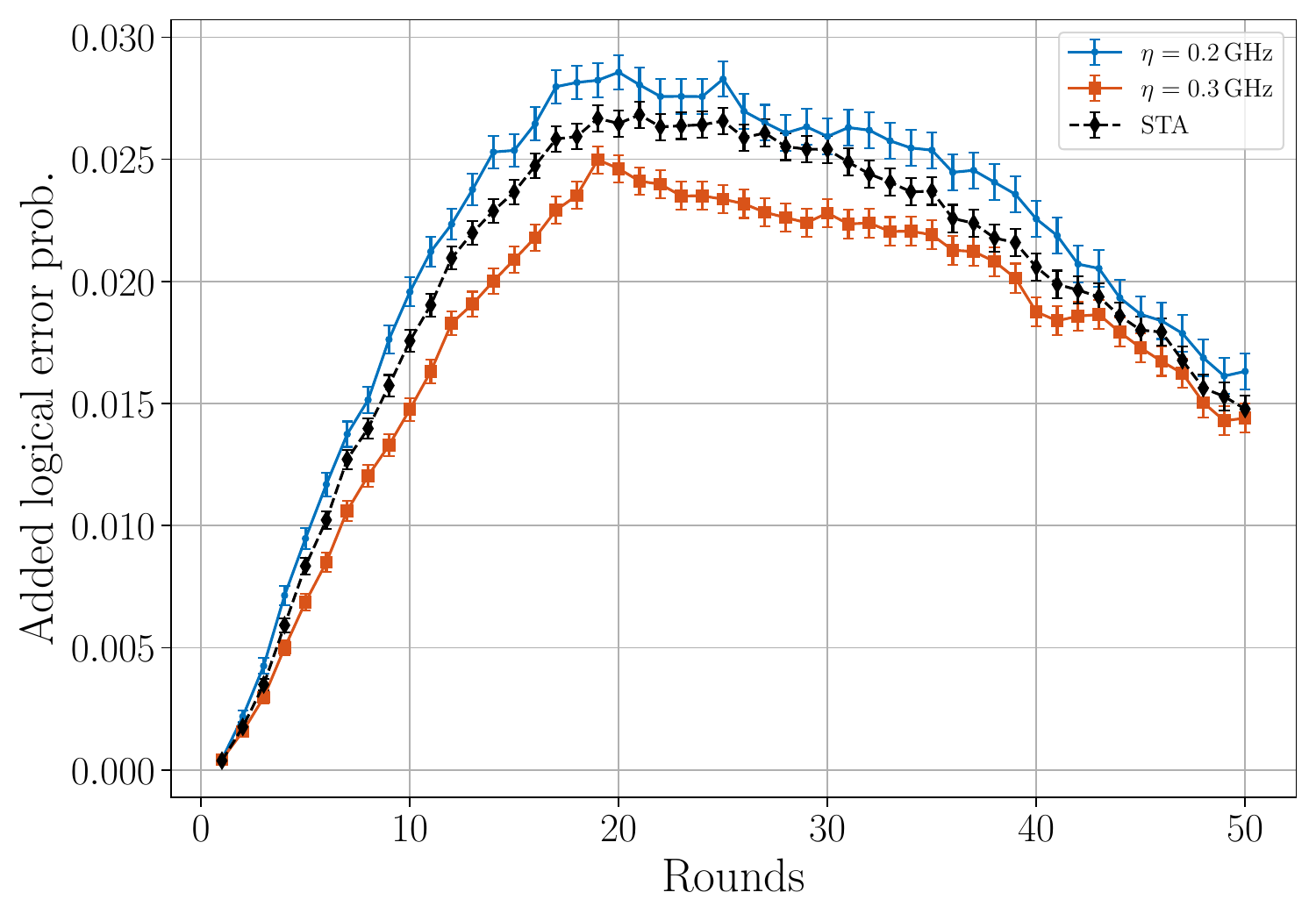}
    \caption{\textbf{Logical error probability for different non-linearities}. We compare the logical error added by leakage for two values of non-linearity $\eta$, and for the STA, which is agnostic of this parameter. The leakage model is that from the main text Fig.~\ref{fig:cz-logical-err}. Data over at least 450k trajectories.}
    \label{fig:eta_LEP}
\end{figure}

\begin{figure*}
    \centering
    \includegraphics[width=1.75\columnwidth]{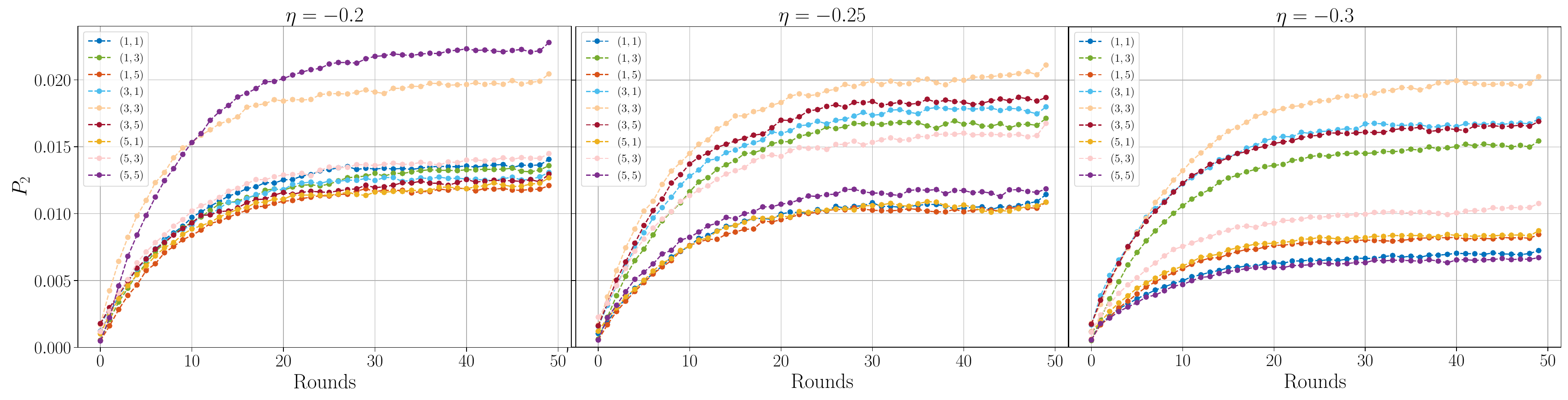}
    \caption{\textbf{Leakage populations with varying non-linearity.} For the distance 3 surface code with strong CZ leakage (cf.~Fig.~\ref{fig:cz-logical-err}), we plot the two state population for all data qubits, where from left to right the non-linearity is $\eta=0.2, 0.25, 0.3$GHz. Data over at least 450k trajectories.}
    \label{fig:leakage-eta}
\end{figure*}

First we demonstrate in Fig.~\ref{fig:leakage-transition-phase} that the transition phase, $\phi_{11,02}$, has little affect on leakage populations. As such, we can justify setting this phase to 0 in the main text simulations.
In contrast however, in Figs.~\ref{fig:eta_LEP}, \ref{fig:leakage-eta} we see that the qubit non-linearity does strongly affect the logical error rate and leakage population. Since the STA is invariant under changes to the non-linearity (by definition), this can lead to discrepancies in logical error properties (as we saw in Fig.~\ref{fig:cz-logical-err}).

\begin{figure*}
    \centering
    \includegraphics[width=1.75\columnwidth]{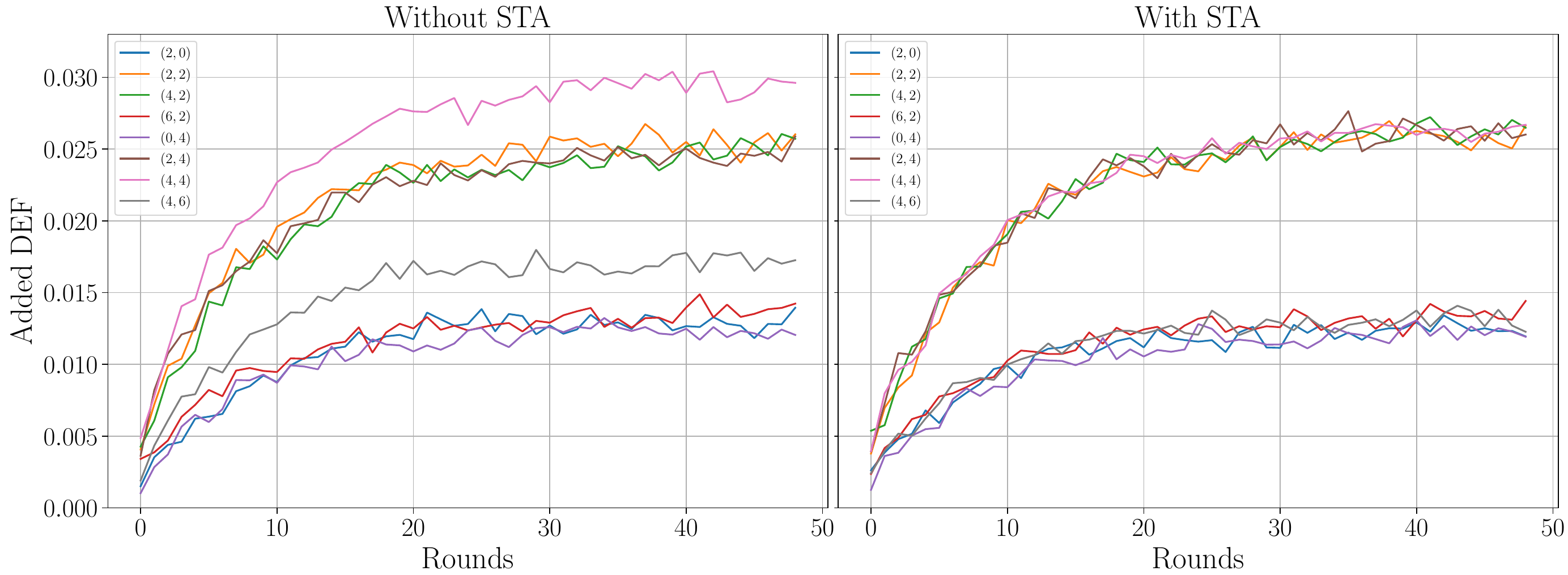}
    \caption{\textbf{Detection event fraction under STA for CZ leakage.} For the distance 3 surface code with strong CZ leakage (cf.~Fig.~\ref{fig:cz-logical-err}), we plot leakage added DEF. The left panel is the full qutrit simulation, and the right panel employs the STA. Here the non-linearity $\eta=0.2$GHz, and the transition phase $\phi_{11,02}=0$. Data is shown for all measure qubits (see Fig.~\ref{fig:SC_schematic}), over at least 450k trajectories.}
    \label{fig:leakage-def}
\end{figure*}

In Fig.~\ref{fig:leakage-def} we study the leakage added DEFs. Again, notice that the STA (right panel) appears to wash out certain coherent effects. In particular, the STA groups all data in two classes; the four central measure qubits (top four curves), and the four boundary qubits (lower four curves). The asymmetry in the left panel is therefore due to coherent effects not captured by the STA. 

We emphasize here again that this model of strong CZ leakage overestimates coherent effects expected from recent experiments (e.g. \cite{scaling_SC_Google_2022}). Below we show that for more physically chosen parameters, that the STA does more accurately reproduce the statistics of interest.

\section{STA with physical error model \label{sec:STA-physical}}
Here we show results pertaining to the repetition/surface codes performance under the STA, where we use approximately physical noise parameters.

In particular we consider a noise model with two forms of leakage, i) from diabatic CZ gates which introduce transitions $|11\rangle \rightarrow \sqrt{1-p}\ket{11} + e^{i\phi_{11,02}}\sqrt{p}\ket{02}$ with probability $p  = 4\times 10^{-4}$ and we set $\phi_{11,02}=0$ (see Fig.~\ref{fig:leakage-transition-phase}), and ii) a thermal heating model with characteristic time-scale (governing $\ket{1}\rightarrow \ket{2}$) $500\mu$s. 
For the CZ leakage, we assume the worst case for the code, where the data qubits leak (since measure qubits are reset to $\ket{0}$ each round after measurement).
The characteristic leakage time-scale (governing $\ket{2} \rightarrow \ket{1}$ transitions) is $10\mu$s.
Additionally the model we use has a $T_1$ time (driving $\ket{1}\rightarrow \ket{0}$ transitions) of $20\mu$s, and a dephasing time-scale $T_{\phi}=40\mu$s. The qutrit non-linearity is 200MHz. We discuss the implementation of these models and show the circuits in App.~\ref{sec:simulation}. Note that each round in our memory experiment takes $1.1\mu$s.
Our choice of parameters above is motivated by recent experimental works, such as \cite{state-preservation-rc-google, scaling_SC_Google_2022, google-leakage-2016}.

\subsection{Repetition code \label{sec:rep-code-STA}}

\begin{figure}
    \centering
    \includegraphics[width=0.98\columnwidth]{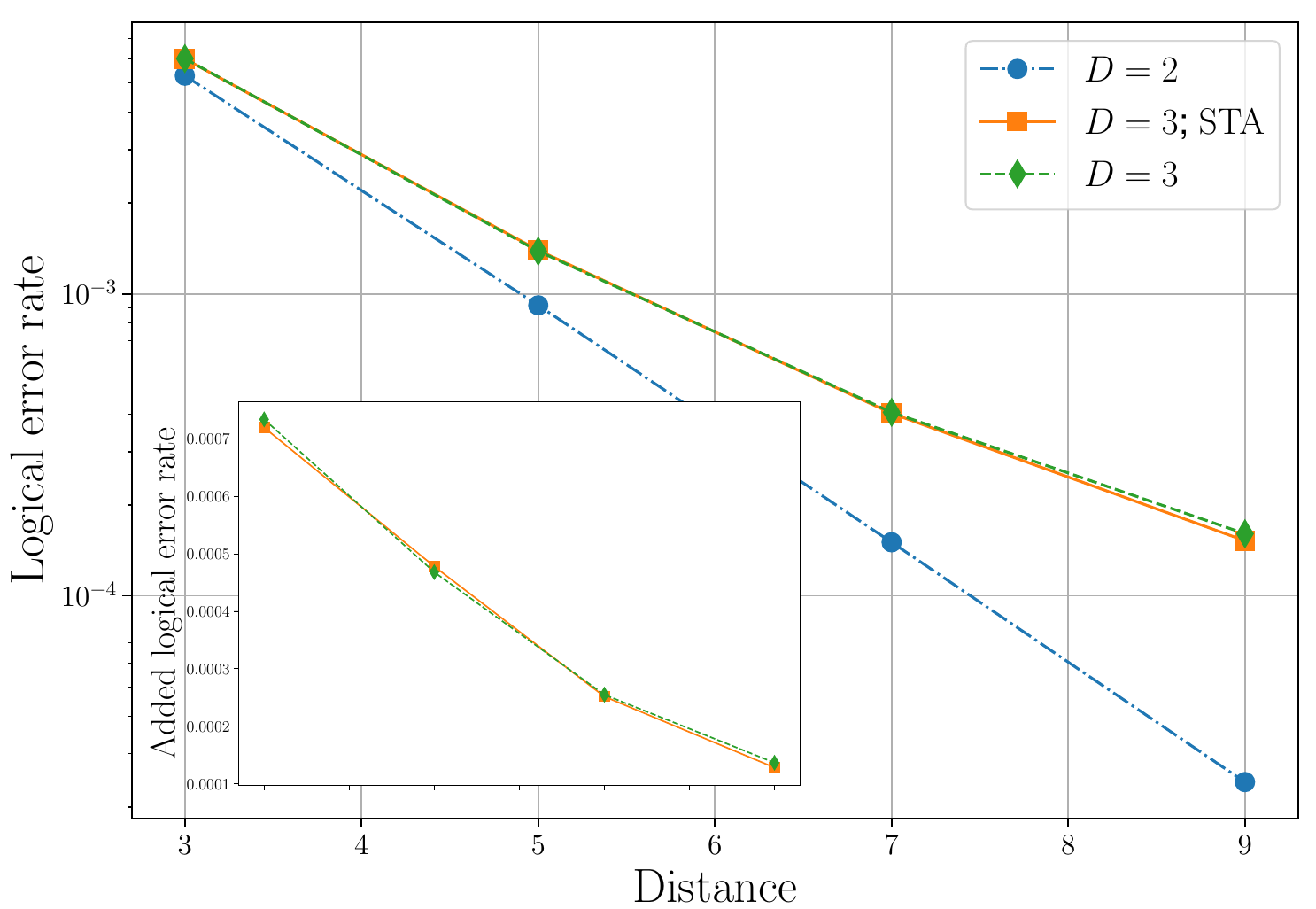}
    \caption{\textbf{Logical error rate for repetition code under STA}. Main figure: Here we compare the extracted logical error rates from a $Z$ basis repetition code memory experiment, in simulations of dimension 2 (blue, circle), simulations including leakage (green, diamond), and simulations with leakage and using the STA (orange, square). Inset: The error rate increase caused by leakage. This is computed by taking the difference of each $D=3$ data point with the corresponding $D=2$ data.
    Simulation parameters are described in the main text. Data is over at least 200k samples per data point.}
    \label{fig:STA_RC}
\end{figure}

Overall with these parameters we find a high quantitative agreement between the STA and the full qutrit simulation. In Fig.~\ref{fig:STA_RC} we study the extracted logical error rate $\epsilon_L$ (see App.~\ref{sec:error-rate}) of the repetition code for distances up to 9.
The logical error rate is a key metric in quantum error correction as it determines how long a quantum memory can be stored.
We see clearly that the STA scales as the full simulation does, with no obvious system-size error dependence. 

\begin{figure}
    \centering
    \includegraphics[width=0.98\columnwidth]{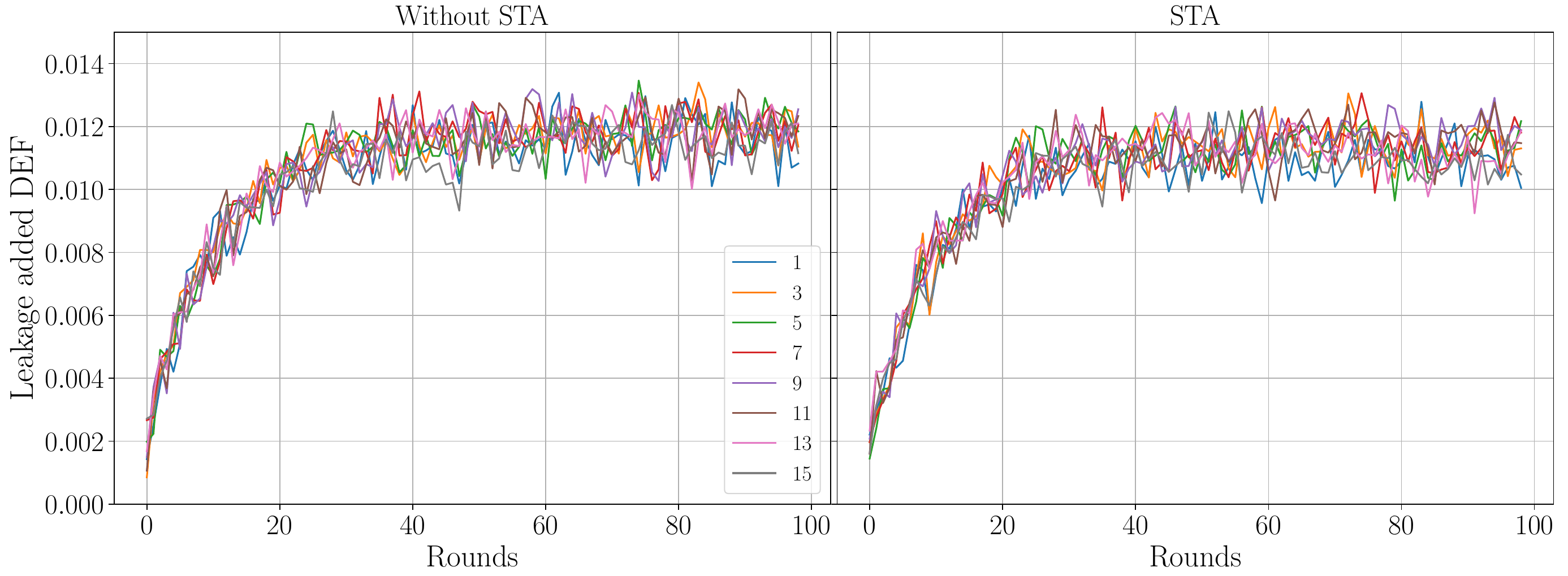}
    \caption{\textbf{Detection event fraction (DEF) of distance 9 repetition code.} The probability of a detection event, per stabilizer (measure qubit) per round. Left shows the full qutrit simulation and right invoking the STA. The long time average for the left plot is 1.18\% vs. 1.13\% in the right plot with STA.
    Data here is over around 225k samples each.}
    \label{fig:DEF_RC9}
\end{figure}

We can additionally consider more `microscopic' properties, such as the distribution of detection events. This is a useful quantity since it is directly related to the noise model implemented.
As a reminder, we call the detection event fraction (DEF) the probability a particular stabilizer measurement registers a detection event, in a given round, where a detection event means the stabilizer value changes from one round to the next. In Fig.~\ref{fig:DEF_RC9} we observe in the distance 9 repetition code that both with and without the STA the DEF added by leakage is close to 1.2\%.

\subsection{Surface code \label{sec:SC-STA}}
\begin{figure}
    \centering
    \includegraphics[width=0.98\columnwidth]{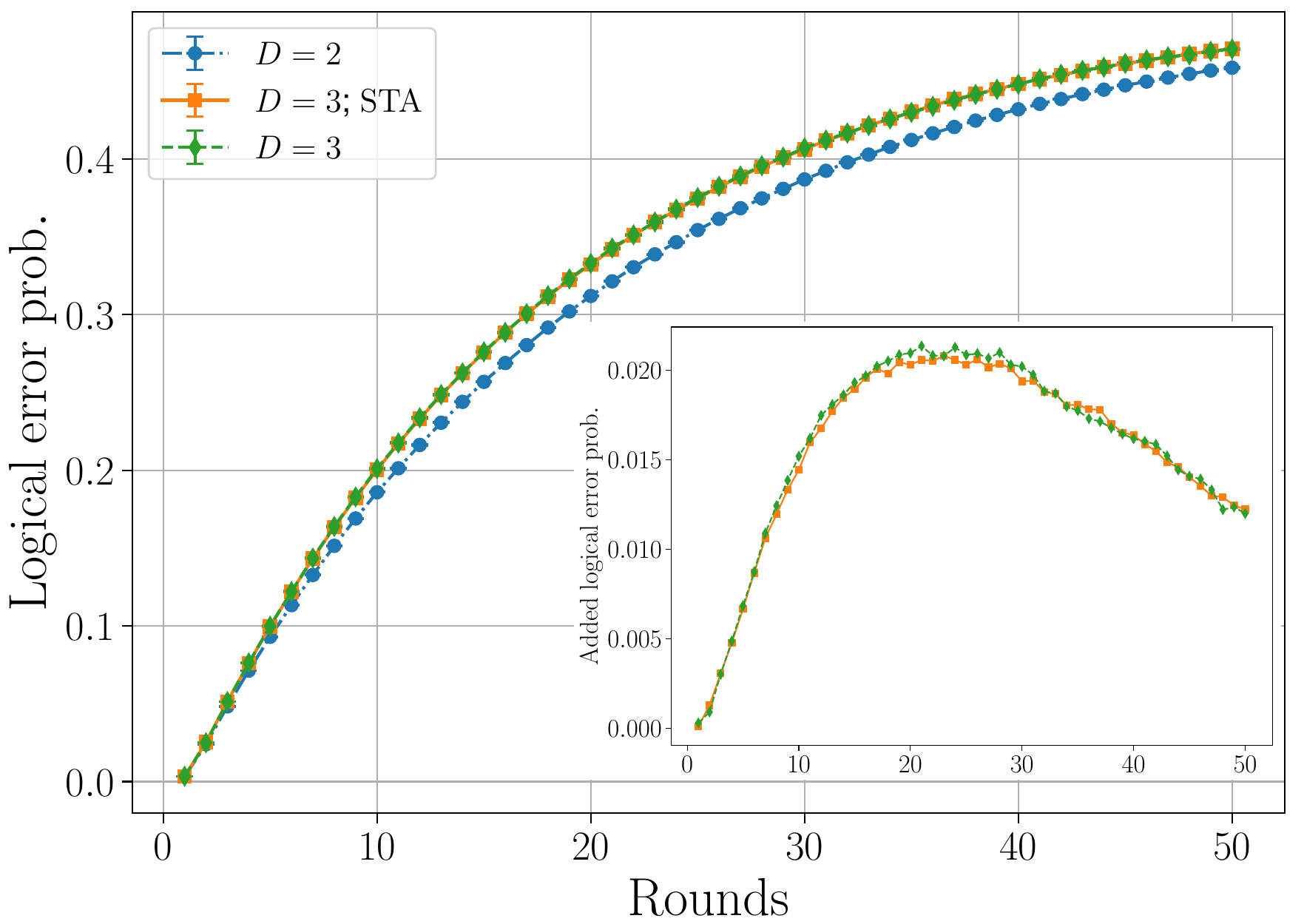}
    \caption{\textbf{Logical error of distance 3 surface code under the STA}. Main figure: The logical error probability for a model with realistic noise parameters, for dimension $D=2$ (blue, circles), and for dimension $D=3$ qutrits (square and diamond). The data of the orange squares corresponds to making the STA. Inset: The added logical error probability by leakage (i.e. the difference between the two curves and the dimension $D=2$ data). Data is from over 1.5M (0.7M) samples with(out) the STA.  Error bars represent the standard error.}
    \label{fig:STA_SC3}
\end{figure}

\begin{figure}
    \centering
    \includegraphics[width=0.98\columnwidth]{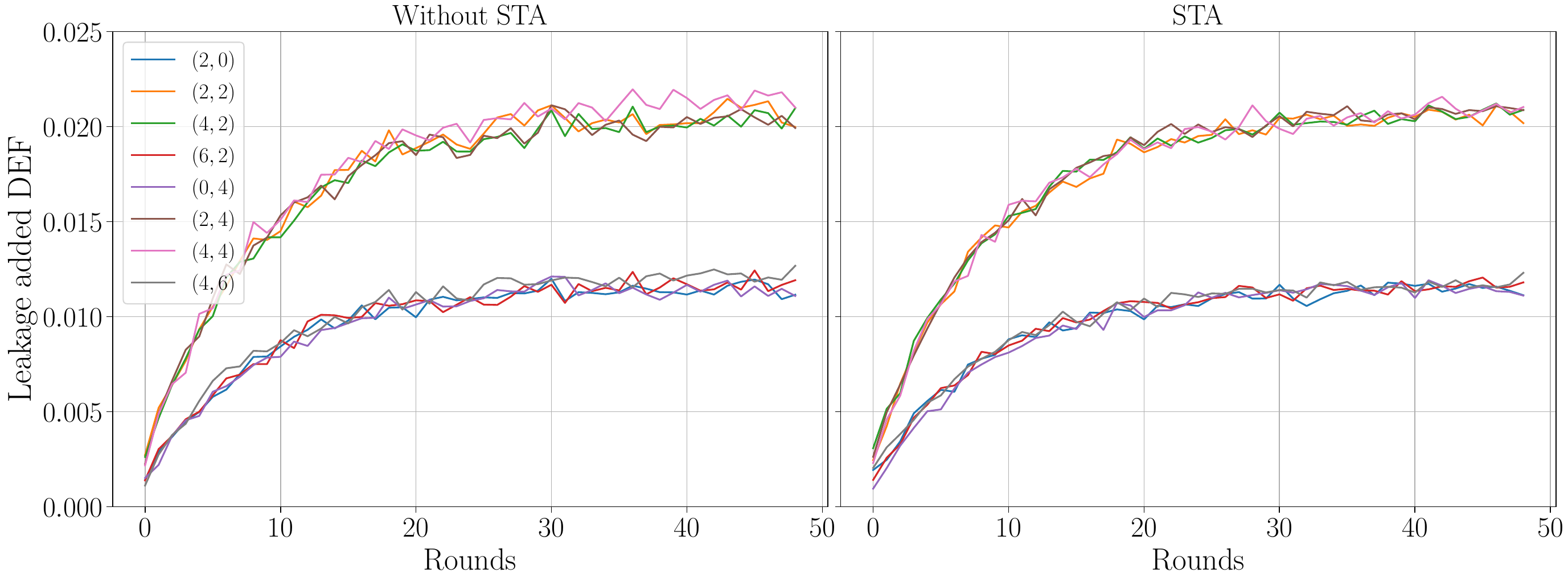}
    \caption{\textbf{Detection event fraction (DEF) of distance 3 surface code with and without STA}. Left shows the detection event fraction added by leakage, per stabilizer (see legend) per round for a full qutrit simulation. The right is the same but employing the STA. Note the two curve groupings correspond to the central stabilizers (which undergo 4 CZ gates per round) versus the boundary (flap) stabilizers (2 CZ per round). The long time mean for the former is 2.06\% in both cases, and the latter 1.15\% (1.16\%) with(out) the STA. The data used is the same as Fig.~\ref{fig:STA_SC3}.}
    \label{fig:STA_DEF_SC3}
\end{figure}

Next we consider a rotated surface code memory experiment at distance 3.
In Fig.~\ref{fig:STA_SC3} we see that broadly the logical error probabilities with and without the STA match well with at most 4\% percentage error during the intermediate rounds which can likely be explained to a large extent by sampling errors. The extracted LERs are 0.0248, 0.0283, 0.0284 for the cases without leakage, with leakage, and with leakage using the STA.
In Fig.~\ref{fig:STA_DEF_SC3} we study the DEF. Here we see the STA reproduces the desired results with remarkably high accuracy; in both cases we observe around a 1.15\% and 2.05\% increase in the number of detection events under leakage, for the boundary and central stabilizers respectively.

\section{Classical rate equation \label{sec:classical-rate}}
Here we discuss two approaches to modelling leakage populations over time.

\subsection{Thermal leakage}
In the case where leakage arises form the thermal heating discussed in App.~\ref{sec:simulation}, and Eq.~\eqref{eq:lindblad_ops}), we can produce a simple classical model of leakage accumulation.
We first assume the 0 and 1 state have equal populations after each round. This is approximately true (up to the effects of noise) for a stabilizer state in the surface code and in bit-flip the repetition code when $X$ flips are performed each round (see Fig.~\ref{fig:rc_sc_schematic}).
We define the per-round probability for a decay from state $2$ to the computational subspace as $\Gamma_-= T_r/T_L$, where $T_L$ is the expected lifetime of state $2$ and $T_r$ is the duration of an error correction round. 
Since the $1\rightarrow 2$ transition time is defined as $T_h/2$ (see App.~\ref{sec:simulation}), the per-round excitation probability to state $2$ is $\Gamma_+ = \frac{1}{2}\frac{2T_r}{T_h} = T_r/T_h$, with $T_h\gg T_L$ (the factor of $1/2$ in front is by the assumption that $P_0\approx P_1$, so heating from the computational subspace occurs at half the rate; we do not consider $0\rightarrow 2$ transitions). 
Letting $P_c(k)$ ($P_2(k)$) denote the probability of a data qubit being in the computational (leaked) subspace at round $k$, we have the following difference equation:
\begin{equation}
\begin{split}
 P_2(k+1) - P_2(k) & = -\Gamma_- P_2(k) + \Gamma_+ P_c(k)  \\ 
& = - P_2(k) \Gamma + \Gamma_+\,,
\end{split}
\label{eq:difference}
\end{equation}
where $\Gamma=\Gamma_++\Gamma_-$.
Assuming boundary condition $P_2(0)=0$, the solution is
\begin{equation}
    P_2(k) = \frac{\Gamma_+}{ \Gamma} (1 - (1-\Gamma)^k)\,.
\end{equation}
 We see the steady state solution is
\begin{equation}
    P_2^{(\infty)} = \frac{\Gamma_+}{\Gamma} \approx \frac{\Gamma_+}{\Gamma_-} = \frac{T_L}{T_h}.
    \label{eq:P2_inf}
\end{equation}
In the case of coherent leakage, this model may not yield accurate results, as it does not take into account any qubit dynamics during the round. This is more properly accounted for below.

\subsection{Coherent leakage}
In the case leakage is coherent in origin (e.g.~via the CZ gate, see Eq.~\eqref{eq:leak_matrix}), we find the above model fails to accurately predict leakage populations. We can, however, construct a more detailed classical three-level Markov process to model leakage. Translating the parameters of this model to the thermal leakage model and applying the STA gives a more accurate match to the coherent leakage populations.

We start with a three level Markov process of the form
\begin{equation}
    \frac{d\vec{P}}{dt} = R \vec{P};\; 
    R = \left(
    \begin{array}{ccc}
        0 & \gamma_{01} & 0 \\
        0 & -(\gamma_{01} + \gamma_{21}) & \gamma_{12}
 \\
 0 & \gamma_{21} & -\gamma_{12}
 \end{array}
    \right)
\end{equation}
with solution $\vec{P}(t) = e^{R t} \vec{P}_0$, where $(\vec{P})_i$ represents the mean population in state $\ket{i}$ ($i\in \{0,1,2\}$) after time $t$. The transition rates (per unit time) $\gamma_{ji}$ correspond to the transitions $|i\rangle \rightarrow |j\rangle$. The related `thermal' model is defined by Lindblad operators with these rates \eqref{eq:lindblad_ops}, where $\gamma_{01}=1/{T_1}$ and $\gamma_{12}=1/{T_L}$ is related to the leakage time-scale $T_L$. In contrast to Eq.~\eqref{eq:difference}, this model takes into account population dynamics between measurements.

In the context of the surface code, the stabilizer measurements at the end of each round results in projecting the qubit subspace to a stabilizer state. The probabilities of measuring a data qubit in state 0 or 1 are then nominally equal, while the 2 state probability is unchanged by the stabilizer measurements. As such, we can model this by a transition matrix
\begin{equation}
    M = \left(
    \begin{array}{ccc}
        1/2 & 1/2 & 0 \\
        1/2 & 1/2 & 0
 \\
 0 & 0 & 1
 \end{array}
    \right).
\end{equation}
Assuming the initial state of each data qubit is $\ket{0}$, the population dynamics over $k$ error correction is then approximated by
\begin{equation}
    \vec{P}_k = (M e^{R T_r})^k \left( \begin{array}{c}
        1 \\ 0 \\ 0 \end{array} \right)\,,
        \label{eq:thermal_approx}
\end{equation}
where $T_r$ is the duration of a single round.

 We can fit the model in Eq.~\eqref{eq:thermal_approx} to the leakage population data over time. Assuming $\gamma_{01}$ and  $\gamma_{12}$ are set by the physical decay times, the model has one unspecified parameter, $\gamma_{21}$, governing the leakage transition $|1\rangle \rightarrow |2\rangle$. The rate extracted by the fit can then be used to construct an effective thermal model, as in the `Thermal Approximation' in Fig.~\ref{fig:cz-logical-err}A, for which the STA is an accurate approximation.

\end{document}